\documentclass[osajnl,onecolumn,showpacs,superscriptaddress,10pt]{revtex4-1} %% use 11pt for Applied Optics
\usepackage{amsmath,amssymb,graphicx}
\usepackage{array} % for better arrays (eg matrices) in maths
\usepackage{astro_bib_macro}
\newcommand{\vecr}{\pmb{r}}
\newcommand{\vs}{\pmb{s}}
\newcommand{\ro}{\pmb{\rho}}

\newcommand{\kap}{\pmb{\kappa}}
\newcommand{\veta}{\pmb{\eta}}

\newcommand{\valpha}{\pmb{\alpha}}
\newcommand{\vphi}{\pmb{\phi}}
\newcommand{\vv}{\pmb{v}}
\newcommand{\vepsilon}{\pmb{\epsilon}}
\newcommand{\lsim} {\buildrel < \over {_\sim}}
\newcommand{\gsim} {\buildrel > \over {_\sim}}

\begin{document}

\title{Laser Tomography Adaptive Optics (LTAO):\\A performance study.}

%% For REVTeX it is possible to automate superscript and e-mail callouts with the superscriptaddress option; see REVTeX4 documentation.

%% \author{Eric Tatulli,$^{1,*}$, A. N. Ramaprakash$^{1}$}
%% \address{$^1$Inter-University Centre for Astronomy and Astrophysics, Ganeshkhind, Pune 411 007, India}
%% \address{$^*$Corresponding author: tatulli@iucaa.ernet.in}

\author{Eric Tatulli}\email{Corresponding author: tatulli@iucaa.ernet.in}
\author{A. N. Ramaprakash}
\affiliation{Inter-University Centre for Astronomy and Astrophysics, Ganeshkhind, Pune 411 007, India}

\begin{abstract} 
We present an analytical derivation of the on-axis performance of Adaptive Optics systems using a given number of guide stars of arbitrary altitude, distributed at arbitrary angular positions in the sky. The expressions of the residual error are given for cases of both  continuous and discrete turbulent atmospheric profiles. Assuming Shack-Hartmann wavefront sensing with circular apertures, we demonstrate that the error is formally described by integrals of products of three Bessel functions. We compare the performance of Adaptive Optics correction when using natural, Sodium or Rayleigh laser guide stars. For small diameter class telescopes ($\lsim 5$m), we show that a few number of Rayleigh beacons can provide similar performance to that of a single Sodium laser, for a lower overall cost of the instrument. For bigger apertures, using Rayleigh stars may not be such a suitable alternative because of the too severe cone effect that drastically degrades the quality of the correction.
\end{abstract}

% REPLACE WITH CORRECT OCIS CODES FOR YOUR ARTICLE
                          % NOTE: \ocis{} IS ALIASED TO \pacs{} BUT MUST
                          % FORMAT THE TERMS CORRECTLY FOR EACH JOURNAL

\maketitle %% null function with osajnl.sty

%%bibtex entry:
%\bibliographystyle{osajnl}
%\bibliography{mybib}
%

\section{Introduction}
The concept of using artificial laser guide stars (LGS) for Adaptive Optics (AO) systems \cite{foy_1,fugate_1}  has been proposed to increase sky coverage by enabling the partial correction of the effects of the atmospheric turbulence in regions where no bright natural guide stars are present in the vicinity of the astrophysical source of interest. In such an instrumental configuration the fundamental limits preventing a perfect correction of the incoming corrugated wavefront have three origins, the latter being specific to the use of LGS:
\begin{enumerate}
\item the inability of the wavefront sensor (WFS) to probe and/or the deformable mirror (DM) to correct some (often the high) frequency components of the turbulent wavefront (the so called \textit{fitting error}), together with the presence of photon and detector noises associated with the WFS measurements. 
\item the spatial and temporal decorrelation between the science and guide star wavefronts, when the guide star is located off-axis and when the correction is applied with a temporal delay due to the finite temporal frequency of the AO control loop. 
\item the spherical nature of LGS wavefront because of the finite altitude of the artificial spot that drives its cone-shaped beam to cross only a fraction of the turbulence seen by the science target, resulting in a additional term in the error budget known as \textit{focus anisoplanatism} \cite{fried_3} or most commonly described as the cone effect \cite{foy_2}.     
\end{enumerate} 
In order to cancel the latter effect that severely reduces the performance of AO systems, it has been proposed to simultaneously use several LGS located at different angular positions in the sky and to perform a 3D mapping of the turbulent volume \cite{foy_2}. For this so-called Laser Tomography Adaptive Optics (LTAO) technique \cite{hubin_1}, each LGS is associated to a dedicated wavefront sensor, and the corrugated wavefront estimated from the 3D-mapped turbulence is compensated with a single deformable mirror (DM) conjugated to the telescope pupil, thus providing a potentially important correction of the atmospheric effects but over a narrow field of view. Generating artificial spots in the sky can be achieved either by Rayleigh back-scattering for low altitude atmospheric layers ($\lsim 20$km) or by excitation of Sodium atoms in the mesospheric Sodium layer located at $\simeq 90$km. Although the first solution requires only mainstream -- hence economical -- laser technology over a large range of wavelengths \cite{fugate_2}, making use of such Rayleigh stars has been mostly abandoned for their low altitude prevents from a good correction of the cone effect, especially for large apertures \cite{lelouarn_1}. Considering its low cost, the potential of Rayleigh LGS however deserves to be quantified in perspective of the financial benefits. On the contrary Sodium stars are often preferred because of the less severe cone effect. It however necessitates custom-made state-of-the-art expensive lasers \cite{bonaccini_1} that drastically increase the budget of the AO system, all the more since several LGSs are contemplated. The capabilities of LTAO technique has been investigated through bench demonstrators \cite{costille_1}  and by means of performance simulations for specific AO systems on large aperture (GALACSI-VLT \cite{strobele_1}, GMT \cite{conanr_1}, ATLAS-ELT \cite{fusco_4}), but no generic theoretical study has been published so far.\\ 
The aim of our paper is thus twofold: in the first part, we provide  in Section (\ref{sec_background}) and (\ref{sec_formalism}) a formal derivation of the performance of LTAO, taking into account in a unified modelling the effects of focus anisoplanatism, incomplete wavefront sensing as well as spatial and temporal decorrelation between the science and guide stars wavefronts, for both continuous and discrete profiles of turbulence. In the second part, we use this analytical framework to quantitatively study  in Sections (\ref{sec_onelgs}) and (\ref{sec_tomo}) the cases of AO systems using one or several LGSs. We finally presents in Section (\ref{sec_compare})  a comparison of the performance that can be expected when using Sodium or Rayleigh lasers with different existing AO systems on telescopes with apertures ranging from $3$m to $10$m.

\section{Background formalism and underlying assumptions} \label{sec_background}
\subsection{Wave propagation and Bessel functions}\label{sec_hj}
Integrals involving the product of Bessel functions have been shown to be an important feature of electromagnetic field propagation through atmosphere \cite{hu_1, sasiela_2}. Following the notation of Hu \textit{et al.} \cite{hu_1}, we introduce the definition of the functions $H2J$ and $H3J$  that will be convenient to express the results of our analytical derivations: 
\begin{eqnarray}
H2J(s,n_1,n_2,a,b) & = & \int_0^{\infty} x^{-s} J^2_{n_1}(ax)J_{n_2}(bx) \mathrm{d}x \\
H3J(s,n_1,n_2,n_3,a,b,c) & = & \int_0^{\infty} x^{-s} J_{n_1}(ax)J_{n_2}(bx)J_{n_3}(cx) \mathrm{d}x
\end{eqnarray}
where $J_{n_1,n_2,n_3}$ are Bessel functions of the first kind of order $n_1$, $n_2$, $n_3$ respectively, and $s,a,b,c$ are parameters of $H2J$ and $H3J$ functions. Integrals of that form are related to Mellin Transform \cite{sasiela_1} and formal evaluations involving gamma and hypergeometric functions \cite{abramowitz_1} can be performed in some specific cases, as provided by Gradshteyn \textit{et al.}\cite{gradshteyn_1} (see Eq. 6.578 $\sharp$1) and by Tyler \cite{tyler_1}. 

\subsection{Independent tip/tilt correction}
Wavefront sensing with monochromatic LGSs is unable to measure the random shift (tip/tilt) of the image because of inverse return of light principle. Several concepts have been proposed to solve this indeterminacy such as making simultaneous use of two small auxiliary telescopes \cite{ragazzoni_3} or two LGS \cite{ragazzoni_2}, by taking advantage of the properties of polychromatic LGS \cite{schock_1} or by adding to the whole AO system a specific instrument dedicated to the estimation of the image displacement by pointing a nearby natural guide star \cite{viard_1}. Since our analysis focuses on performance of LGS Adaptive Optics, that is the correction of higher order modes than tip and tilt, we assume in the following that these are estimated independently and fully corrected. In order to take into account partial tip/tilt correction, a quadratic error must be added to the error budget following e.g. the formalism of D. Sandler \cite{sandler_1}  that models the atmospheric tip/tilt error (influence of higher modes on the estimation of tip/tilt \cite{yura_1}), tip/tilt anisoplanatism error and photon/detector noise associated to the tip/tilt measurements.

\subsection{Science star turbulent wavefront}
We define $\Phi(\vecr)$ as the turbulent phase of the plane wavefront arising from the science star. Using Zernike polynomials, the piston/tip-tilt removed science phase can be written as:
\begin{equation}
\Phi(R\ro) = \sum_{j=3}^{\infty}\phi_j Z_j(\ro)\label{eq_zernphi}
\end{equation}
with $R$ the radius of the telescope aperture, and $\ro=\vecr/R$, the polynomials being defined over the unit radius circle. The piston mode is also not considered as it is irrelevant for AO correction and wavefront sensing issues. The statistics of the turbulent science phase is characterized by the covariance matrix $\mathrm{Cov}(\vphi)=<\vphi\vphi^T>$ ,
where $<>$ denotes the statistical average and $^T$ is  the transpose operator. Following Noll description \cite{noll_1}, the turbulence variance $\sigma^2_{\phi}$, that is the trace of the covariance, writes:
\begin{equation}   
\mathrm{Tr}\left\{\mathrm{Cov}(\vphi)\right\} = \sigma^2_{\phi} =  0.135 \left(\frac{D}{r_0}\right)^{\frac{5}{3}} \label{eq_noll}
\end{equation} 
where $D=2R$ is the diameter of the telescope and $r_0$ is the Fried parameter defined at zenith as \cite{fried_1}:
\begin{equation}
r_0 = \left[\frac{0.033(2\pi)^{-\frac{2}{3}} \left(\frac{2\pi}{\lambda}\right)^2}{0.023}\int_0^{\infty}C^2_n(h) \mathrm{d}{h}\right]^{-\frac{3}{5}} 
\end{equation}
$C^2_n(h)$ is the atmospheric structure constant of the refractive index along the altitude $h$ above the telescope.

\subsection{Spherical LGS wavefronts: the cone effect}
\begin{figure*}[htbp]
\begin{center}
\begin{tabular}{cc}
\includegraphics[width=0.50\textwidth]{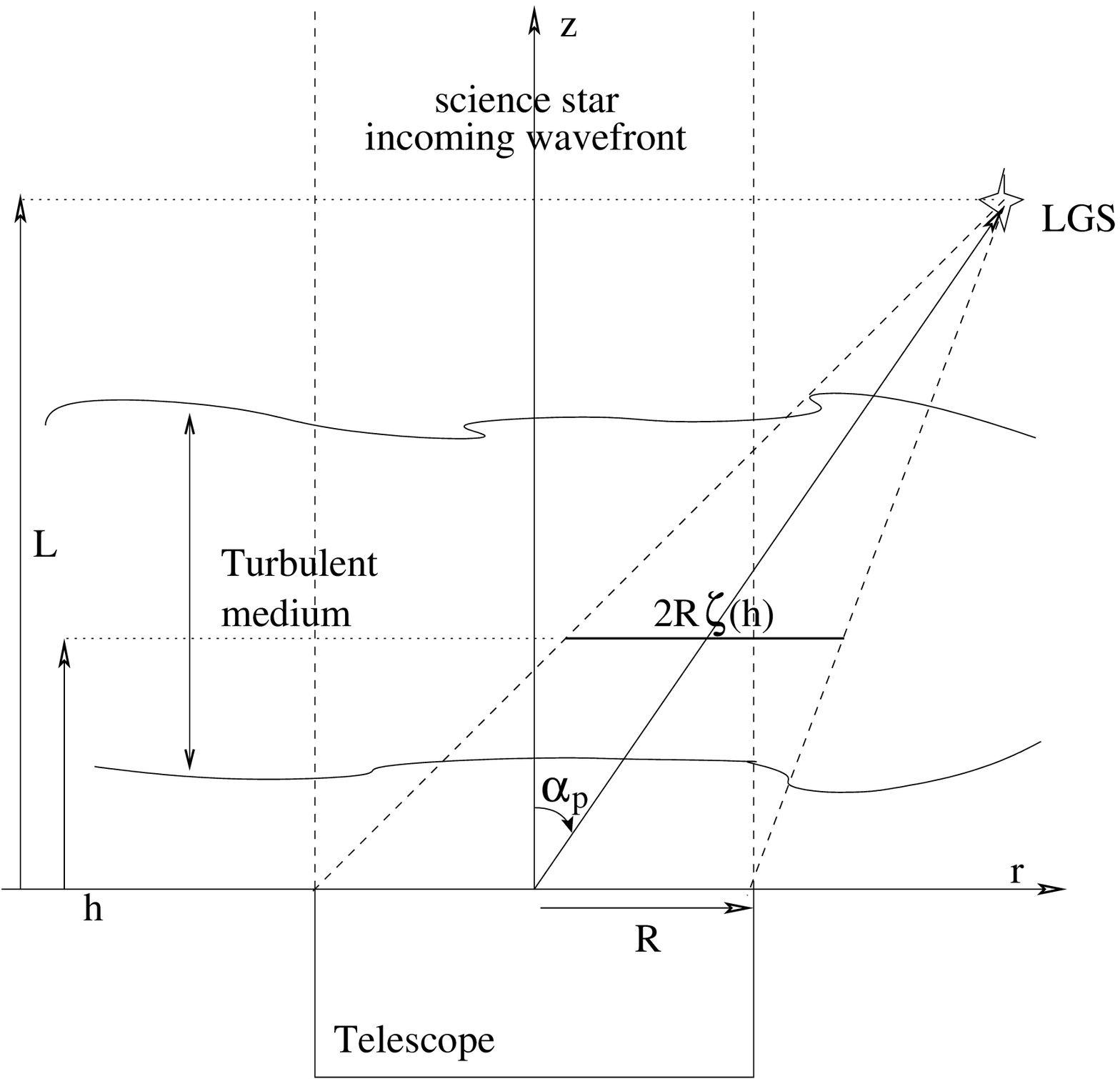}&\includegraphics[width=0.4\textwidth]{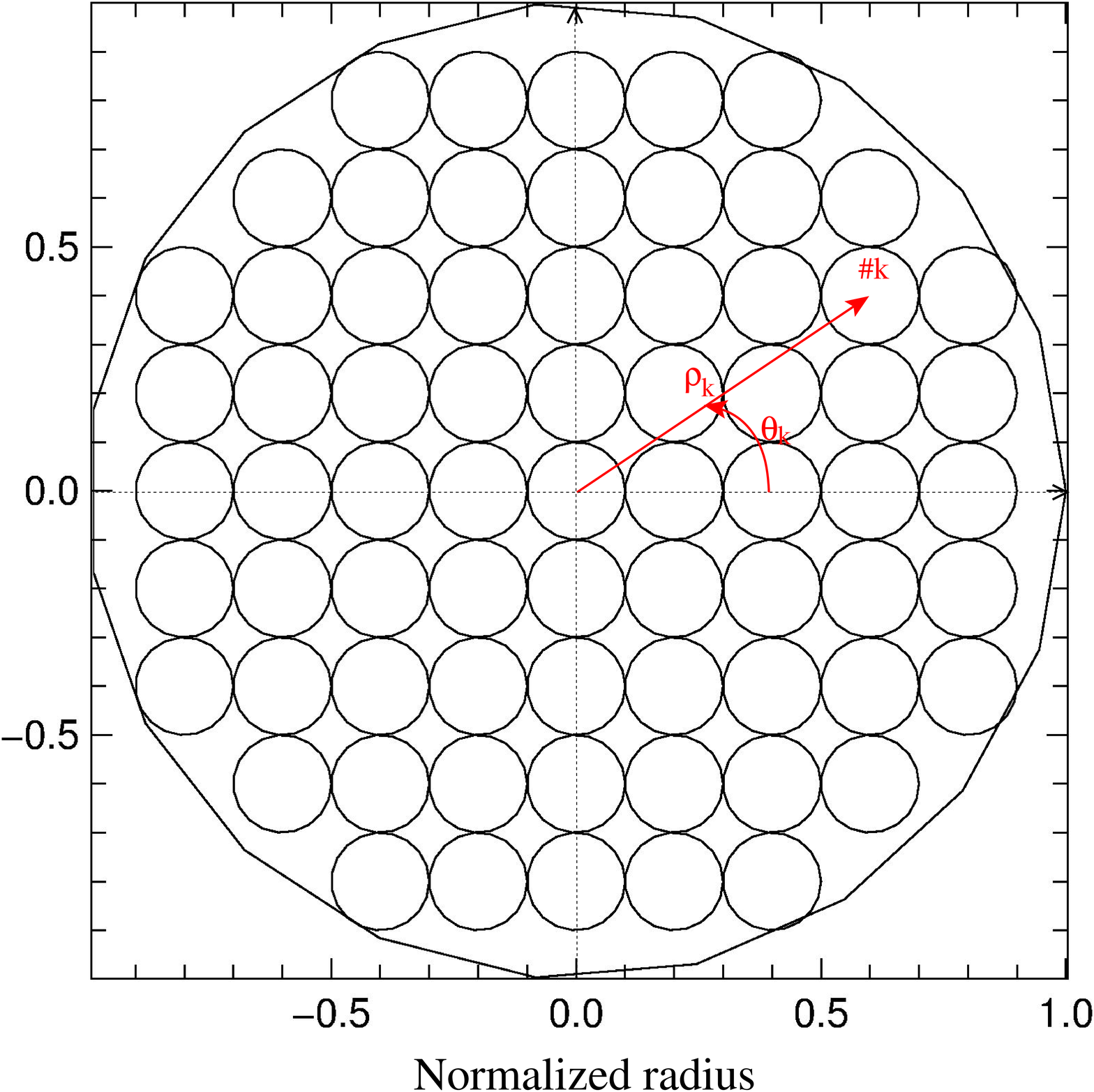}
\end{tabular}
\caption{\label{fig_sketch1} Left: Sketch of LGS AO observations. The angular location of the laser spot is $\valpha_p$, its height is $L$. For a given altitude $h$, the laser beacon light crosses a turbulence portion of radius $\zeta(h)R$, with $\zeta(h)=\frac{L-h}{h}$. Right: Representation of the SH subapertures, in polar coordinates. The $k^{th}$ subaperture is located at a normalized radius $\rho_k$ and an angle $\theta_k$.}
\end{center}
\end{figure*}
We call $\Phi^{lgs}(\vecr,\valpha_p)$ the turbulent phase of the spherical wavefronts coming from the $N_{lgs}$ laser guide stars located at respective angular position $\valpha_p$, $p \in [1..N_{lgs}]$ that are used to probe the atmospheric turbulence. The portion of atmosphere crossed by the laser beams -- therefore the turbulent LGS phase -- depends on $\valpha_p$. The spatial covariance $B_{\Phi}^{lgs}(R\ro)$ of the LGS turbulent phase that characterizes its statistical properties is however independent of this angular location and  can be written as:
\begin{eqnarray}
B_{\Phi}^{lgs}(R\ro)&=& <\Phi^{lgs}(R[\ro_1+\ro], \valpha_p)\Phi^{lgs}(R\ro_1, \valpha_p)>   \\
&=& \left(\frac{2\pi}{\lambda}\right)^2 <\int_0^{L} n(R\zeta(h)[\ro_1+\ro], \valpha_p)\mathrm{d}h \int_0^{L} n(R\zeta(h)\ro_1, \valpha_p) \mathrm{d}h> \nonumber \\
&=& \left(\frac{2\pi}{\lambda}\right)^2 \int_0^{L} B^h_{\Delta{n}}(R\zeta(h)\ro)\mathrm{d}h \label{eq_covlgs}
\end{eqnarray}
where $n$ is the refractive index and $B^h_{\Delta{n}}$ is the covariance of its fluctuation for the turbulent layer located at the altitude $h$ and of infinitesimal thickness $\delta{h}$, and assuming that these layers are statistically independent (small perturbations and near-field approximations \cite{roddier_1}). Due to the spherical nature of the LGS wavefront (cone effect), the fraction of the turbulence $\zeta(h)$ seen by the LGS beam at the altitude $h$ is $\zeta (h) = \left(\frac{L-h}{L}\right)$ with $L$ the altitude of the LGS, as shown in Fig. (\ref{fig_sketch1}, left). \\
The power spectrum $W^{lgs}_{\Phi}(\kap)$ of the LGS phase is by definition the Fourier Transform of its spatial covariance and thanks to Eq. (\ref{eq_covlgs}) can be written as:
\begin{equation}
 W^{lgs}_{\Phi}(\kap)= \left(\frac{2\pi}{\lambda}\right)^2\int_0^{L} \frac{1}{[R\zeta(h)]^2} W^{h}_{\Delta{n}}\left(\frac{\kap}{R\zeta(h)}\right)\mathrm{d}h
\end{equation}
Under Kolmogorov statistics hypothesis \cite{kolmogorov_1,tatarski_1}, the refractive index fluctuation power spectrum $W^{h}_{\Delta{n}}(\kap)$ is given by:
\begin{equation}
W^{h}_{\Delta{n}}(\kap) = 0.033(2\pi)^{-\frac{2}{3}} |\kap|^{-\frac{11}{3}} C^2_n(h) = \left(\frac{\lambda}{2\pi}\right)^2 0.023 r_0^{-\frac{5}{3}} |\kap|^{-\frac{11}{3}} \frac{C^2_n(h)}{\int_0^{\infty} C^2_n(h)  \mathrm{d}h}  \label{eq_wn}
\end{equation}
and $W^{lgs}_{\Phi}(\kap)$ takes the final form:
\begin{equation}
W^{lgs}_{\Phi}(\kap)=0.023 \left(\frac{R}{r_0}\right)^{\frac{5}{3}}|\kap|^{-\frac{11}{3}}\frac{\int_0^{L} [\zeta(h)]^{\frac{5}{3}} C^2_n(h)\mathrm{d}h }{\int_0^{\infty} C^2_n(h)  \mathrm{d}h}
\end{equation}
The ratio of the integrals over the altitude captures the cone effect due to the finite altitude of the LGS. In case of a plane wavefront we have $L=\infty$ and $\zeta(h)=1$, hence the ratio is equal to one and we obtain the definition of the classical Kolmogorov phase power spectrum. \\
Finally we describe the LGS phase over the Zernike polynomial basis as following:
\begin{equation}
\Phi^{lgs}(R\ro, \valpha_p) = \sum_{j=1}^{\infty}\phi^{lgs}_j(\valpha_p) Z_j(\ro)\label{eq_zernphilgs}
\end{equation}

\subsection{Control loop delay}
An AO loop works at a finite speed (roughly a few hundred Hz), which translates into a time delay $\tau$ between the observation of the scientific source and the actual correction of the atmospheric perturbations from the guide star. In such a case, the science star turbulent phase $\Phi(\vecr,t)$ taken at given time $t$ will be corrected from the LGS phase $\Phi^{lgs}(\vecr,\valpha_p, t+\tau)$ taken at a time $t+\tau$. Under Taylor hypothesis of ``frozen turbulence''  this time delay can be transformed into a spatial shift $\Delta\ro =\tau \vv(h)$, where $\vv(h)$ is the wind speed vector for the altitude $h$. The crossed-covariance between the science star and guide star phases can thus be computed as:
\begin{eqnarray}
B_{\Phi^{lgs}}^{\Phi}(R\ro, R\ro_1, \valpha_p, \tau)&=& <\Phi(R[\ro_1+\ro],t)\Phi^{lgs}(R\ro_1, \valpha_p, t+\tau)>   \\
&=& \left(\frac{2\pi}{\lambda}\right)^2 <\int_0^{L} n(R[\ro_1+\ro])\mathrm{d}h \int_0^{L} n(R\zeta(h)\ro_1+h\valpha_p+\tau \vv(h)) \mathrm{d}h> \nonumber \\
&=& \left(\frac{2\pi}{\lambda}\right)^2 \int_0^{L} B^h_{\Delta{n}}(R\ro_1[1-\zeta(h)]+R\ro-h\valpha_p-\tau \vv(h))\mathrm{d}h \label{eq_crosscovlgs}
\end{eqnarray}
We emphazise that, at the difference of the plane and spherical wavefront phase covariances, the cross-covariance is a non-stationnary process since it depends on the location $\ro_1$ where this quantity is computed from.\\
For describing the wind associated to the turbulent layers, Bufton \cite{bufton_1} has provided an empirical law for the wind speed modulus:
\begin{equation}
v(h) = 5 + 30\exp\left[-\frac{(h-9.4)^2}{4.8^2}\right]
\end{equation}
where the numbers outside the brackets are in meters per second. From the wind speed average $\overline{v}$, one can estimate the coherence time  of the turbulence $t_0$, using the definition of Greenwood  \cite{greenwood_1}:
\begin{equation}
t_0 = 0.314 \frac{r_0}{\overline{v}}
\end{equation}

\subsection{Wavefront sensing}
We assume that identical Shack-Hartmann (SH) wavefront sensors \cite{rousset_1} are associated to every LGS beam. We call $M_s$ the number of subapertures of each SH that will therefore provide $2M_s$ slopes measurements corresponding to the LGS turbulent phase. We denote $\vs(\valpha_p) = [\vs^x(\valpha_p), \vs^y(\valpha_p)]$ assuch slopes measurements, in $x$ and $y$ directions. Considering the $k^{th}$ subaperture, the SH provides the derivative of the LGS phase as following \cite{rousset_1}:
\begin{equation}
\vs_k^{x,y}(\valpha_p) = \frac{\lambda}{2\pi A_s} \int_{subap_k} \frac{\partial \Phi^{lgs}(\vecr,\valpha_p)}{\partial{x,y}} \mathrm{d}^2\vecr =  \frac{\lambda{R}}{2\pi A_s} \int_{subap_k} \frac{\partial}{\partial{x,y}} [\Phi^{lgs}(R\ro,\valpha_p)] \mathrm{d}^2\ro \label{eq_alpha}
\end{equation}
$A_s$ is the area of the subaperture, and $\lambda$ the wavelength of the AO WFS path. 
From Eqs. (\ref{eq_alpha}) and (\ref{eq_zernphilgs}), we can introduce the interaction matrix $D_{\infty}$  that converts the LGS phase Zernike coefficients into SH slope measurements:
\begin{equation}
\vs=D_{\infty}\vphi^{lgs}; \label{eq_sh}
\end{equation}
Note that $D_{\infty}$ is block-diagonal, the number of blocks being equal to the number of LGS/AO used. Each block is made of two sub-matrices $[D^{x},D^y]$ that account for the slopes measurements in both directions, that is:
\begin{equation}
 D^{x,y}_{kj}=\frac{\lambda{R}}{2\pi A_s} \int_{subap_k} \frac{\partial Z_j(\ro)}{\partial{x,y}}  \mathrm{d}^2\ro = \frac{\lambda{R}}{2\pi A_s} \int \Pi_s^k\left(\frac{R}{R_s}\ro\right) \frac{\partial Z_j(\ro)}{\partial{x,y}}  \mathrm{d}^2\ro \label{eq_system}
\end{equation}
where $ \Pi_s^k\left(\frac{R}{R_s}\ro\right)$ is a function of the $k^{th}$ subaperture and $R_s$ its characteristic size. It can be rewritten in the form $\Pi_s^k\left(\frac{R}{R_s}\ro\right) = \Pi_s\left(\frac{R}{R_s}[\ro-\ro_k]\right)$, where $\ro_k=[\rho_k, \theta_k]$ is the normalized coordinate vector of the $k^{th}$ subaperture, with respect to the center of the telescope aperture, as shown in Fig (\ref{fig_sketch1},right). For a circular subaperture of radius $R_s$, we have $A_s=\pi{R_s}^2$ and the Fourier Transform $\widehat{\Pi_s^k}(\kap)$ of the  $k^{th}$ subaperture can be written as:
\begin{eqnarray}
\widehat{\Pi_s^k}(\kap) = \int \Pi_s\left(\frac{R}{R_s}[\ro-\ro_k]\right) \exp^{-2i\pi\ro.\kap}\mathrm{d}^2 {\ro} = \left[\frac{R_s}{R}\right] \frac{J_1\left(2\pi\frac{R_s|\kap|}{R}\right)}{|\kap|}\exp^{-2i\pi\ro_k.\kap} \label{eq_ftpis}
\end{eqnarray}
In such a case, the elements of the interaction matrix can be computed formally in terms of integrals of products of three Bessel functions, as demonstrated in App. A.3. Using notations of Sect. \ref{sec_hj}, we have:
\begin{eqnarray}
 D^{x}_{kj} &=& \frac{\lambda}{2\pi R_s} s_{n,m} \left[\beta^{x}_{|m|-1}(\theta_k)H3J(0,1,n+1,|m|-1, R_s/R,1,\rho_k) \right.  \nonumber \\
&& \hskip40pt - \left. \beta^{x}_{|m|+1}(\theta_k)H3J(0,1,n+1,|m|+1, R_s/R,1,\rho_k)\right]
\\
 D^{y}_{kj} &=& \frac{\lambda}{2\pi R_s} s_{n,m} \left[\beta^{y}_{|m|-1}(\theta_k)H3J(0,1,n+1,|m|-1, R_s/R,1,\rho_k) \right.  \nonumber \\
&& \hskip40pt + \left. \beta^{y}_{|m|+1}(\theta_k)H3J(0,1,n+1,|m|+1, R_s/R,1,\rho_k)\right]\end{eqnarray}
where $n$ and $m$ are respectively the radial degree and the azimuthal frequency associated to the $j^{th}$ Zernike polynomial and $s_{n,m}$, $\beta^{x,y}_{|m| \pm 1}$ are defined by:
\begin{eqnarray}
&&s_{n,m} =  i^{|m|}(-1)^\frac{3n}{2}\sqrt{n+1}\left\{\begin{array}{ll} \sqrt{2}&\mathrm{if}~m \ne 0 \\ 1&\mathrm{if}~m=0\end{array}\right.; \label{eq_snm}\\
&&\beta^{x,y}_{|m| \pm 1,k}(\theta_k) = \left\{\begin{array}{lrr} \cos([|m| \pm 1]\theta_k),& -\sin([|m| \pm 1]\theta_k) & \mathrm{if}~m \ge 0 \\ \sin([|m| \pm 1]\theta_k),& \cos([|m| \pm 1]\theta_k) & \mathrm{if}~m<0 \end{array}\right. 
\end{eqnarray}
When using LGS beacons, the SH will not be sensitive to the tip/tilt modes of the LGS phase. As a result the tip and tilt contributions to the slopes must be removed, such that the effective measured slopes $\widehat{\vs}$ are given by:
\begin{eqnarray}
\widehat{\vs^x} = \vs^x -  \left(\frac{\lambda}{\pi{R}}\right) \phi^{lgs}_1,~~\widehat{\vs^y} = \vs^y -  \left(\frac{\lambda}{\pi{R}}\right) \phi^{lgs}_2  \label{eq_nottslope}
\end{eqnarray}
where $ \phi^{lgs}_1$ and $ \phi^{lgs}_2$ are respectively the tip and tilt Zernike coefficients of the LGS turbulent wavefronts. 

\subsection{Perfect deformable mirrors}
For sake of simplicity we assume in the following that the deformable mirror (DM) is able to perfectly reproduce the shape of the wavefront provided by the wavefront sensors. In practice there is however a mismatch between the desired wavefront and the surface that the mirror will eventually take, since the number of actuators that shape the surface of the mirror is not infinite.  This mismatch can be modelled by taking into account the projection of the slopes onto the DM modes, that is the actuators responses.  We refer to the work of Wallner \cite{wallner_1} (single guide star case) and Tokkovinin \textit{et. al.} \cite{tokovinin_1} (multiple guide stars case) for a modelling of the problem that includes this effect.

\subsection{Wavefront reconstruction and residual phase error}
We call $\widetilde{\Phi}(\vecr)$ the estimated turbulent phase from the slope measurements and $\widetilde{\vphi}$ its related Zernike coefficients vector.The residual phase variance is by definition the variance of the phase difference integrated over the pupil of the telescope:
\begin{equation}
\sigma^2_{res} = \int \Pi_p(\ro) <|\Phi(R\ro)-\widetilde{\Phi}(R\ro)|^2> \mathrm{d}^2\ro \label{eq_res_error}
\end{equation}
where $\Pi_p(\ro)$ is the unitary pupil function.\\
The computation of $\widetilde{\Phi}$ from the measurements $\vs$ is a linear fitting process. We introduce $M$ the so-called control matrix \cite{wallner_1} representing this process. We thus can write the following relationship:
\begin{equation}
\widetilde{\vphi} = M (\widehat{\vs} + \vepsilon)
\end{equation}
where $\vepsilon$ is the additive (i.e. photon, detector) noise associated to the slopes. Data cosmetics (flat-field, dark current etc.) are not considered in this paper since these effects are assuemd to be removed through proper calibration. \\
If we assume an aperture without central obstruction, standard Zernike polynomials form an orthonormal basis and equation (\ref{eq_res_error}) simplifies as:
\begin{equation}
\sigma^2_{res} = <\|\vphi-\widetilde{\vphi}\|^2>_{atm,\epsilon} =  <\|\vphi- M (\widehat{\vs} + \vepsilon)\|^2>_{atm,\epsilon}  \label{eq_res_error2}
\end{equation}
where $<>_{atm,\epsilon}$ is the  average over both the atmosphere and the additive noise statistics. The explicit form of $M$ will be investigated in Sect. (\ref{subsec_mmse}).

\section{Computation of the residual phase error}\label{sec_formalism}
The aim of this Section is threefold: first we provide the formal expression of the residual phase error in the general case of multiple LGS AO correction and continous turbulent atmospheric profile. However, performing tomography of the turbulence requires to describe the atmosphere as thin discrete turbulent layers located at specific heights. In this respect, we also provide the computation of the residual error using an independent matrix-oriented approach. From this latter modelling, we finally derive the expression of the optimal control matrix $M$ that enables to minimize the residual error.
\subsection{General analytical approach}\label{sec_gen_approach}
With further hypothesis that atmospheric and additive noises are independent, the matrix expression of previous equation is:
\begin{eqnarray}
\sigma^2_{res} &=& \mathrm{Tr}\left\{<(\vphi- M (\widehat{\vs} + \vepsilon))(\vphi- M (\widehat{\vs} + \vepsilon))^{T}>_{atm,\epsilon}\right\} \nonumber \\
&=& \mathrm{Tr}\left\{<\vphi\vphi^T> + M<\widehat{\vs}\widehat{\vs}^T>M^T-<\vphi\widehat{\vs}^T>M^T-M<\widehat{\vs}\vphi^T> +  M<\vepsilon\vepsilon^T>M^T\right\} \nonumber \\ 
&=& \mathrm{Tr}\left\{\mathrm{Cov}(\vphi) +  M\mathrm{Cov}(\widehat{\vs})M^T - 2\mathrm{Cov}(\vphi,\widehat{\vs})M^T + M\mathrm{Cov}(\vepsilon)M^T\right\} \label{eq_sig2res}
\end{eqnarray}
$\mathrm{Cov}(\vs)$ denotes the covariance of the slopes measurements. As $\widehat{\vs}$ is the concatenation of $x$ and $y$ slopes for each LGS located at $\valpha_p$, the elements of the matrix results in the computation of three  moments $C_s^{xx}$,  $C_s^{yy}$,$C_s^{xy}$ with $C_s^{xx}=<\widehat{s_k^x}(\valpha_p)\widehat{s_l^x}(\valpha_q)>$, $C_s^{yy}=<\widehat{s_k^y}(\valpha_p)\widehat{s_l^y}(\valpha_q)>$ and $C_s^{xy}=<\widehat{s_k^x}(\valpha_p)\widehat{s_l^y}(\valpha_q)>$, that, according to Eq. (\ref{eq_nottslope}), leads to:\\
\begin{eqnarray}
C_s^{xx}&=&<s_k^x(\valpha_p)s_l^x(\valpha_q)>+ \left(\frac{\lambda}{\pi{R}}\right)^2< \phi^{lgs}_1(\valpha_p)\phi^{lgs}_1(\valpha_q)>\nonumber\\
&&-\left(\frac{\lambda}{\pi{R}}\right)\left[<s_k^x(\valpha_p)\phi^{lgs}_1(\valpha_q)>+<s_l^x(\valpha_q)\phi^{lgs}_1(\valpha_p)>\right]\\
C_s^{yy}&=&<s_k^y(\valpha_p)s_l^y(\valpha_q)>+ \left(\frac{\lambda}{\pi{R}}\right)^2< \phi^{lgs}_2(\valpha_p)\phi^{lgs}_2(\valpha_q)>\nonumber\\
&&-\left(\frac{\lambda}{\pi{R}}\right)\left[<s_k^y(\valpha_p)\phi^{lgs}_2(\valpha_q)>+<s_l^y(\valpha_q)\phi^{lgs}_2(\valpha_p)>\right]\\
C_s^{xy}&=&<s_k^x(\valpha_p)s_l^y(\valpha_q)>+ \left(\frac{\lambda}{\pi{R}}\right)^2< \phi^{lgs}_1(\valpha_p)\phi^{lgs}_2(\valpha_q)>\nonumber\\
&&-\left(\frac{\lambda}{\pi{R}}\right)\left[<s_k^x(\valpha_p)\phi^{lgs}_2(\valpha_q)>+<s_l^y(\valpha_q)\phi^{lgs}_1(\valpha_p)>\right]
\end{eqnarray}
The formal expressions of the moments involved in the computation of $\mathrm{Cov}(\vs)$  are given in Appendix B. For the case of SH circular subapertures, the moments can be written using $H2J$ and $H3J$ functions: 
\begin{eqnarray}
\left\{\begin{array}{c} <s_k^x(\valpha_p)s_l^x(\valpha_q)> \\  <s_k^y(\valpha_p)s_l^y(\valpha_q)> \\  <s_k^x(\valpha_p)s_l^y(\valpha_q)> \end{array}\right\} &=& \frac{0.0493}{ \int_0^{\infty} C_n^2(h)\mathrm{d}h}\left(\frac{D}{r_0}\right)^{\frac{5}{3}} \left(\frac{\lambda}{R_s}\right)^2\nonumber \\
&\times& \int_0^L \mathrm{d}h~[\zeta(h)]^{\frac{5}{3}.}C_n^2(h)~[\left\{\begin{array}{r}1 \\ 1 \\ 0\end{array}\right\}HJ2(8/3,1,0,R_s/R, \rho^{pq}_{kl}(h)) \nonumber \\ 
&& - \left\{\begin{array}{r} \cos(2\theta^{pq}_{kl}(h)) \\ -\cos(2\theta^{pq}_{kl}(h)) \\ \sin(2\theta^{pq}_{kl}(h))\end{array}\right\}  HJ2(8/3,1,2,R_s/R, \rho^{pq}_{kl}(h))] 
\end{eqnarray}
\begin{eqnarray}
\left\{\begin{array}{c} <\phi^{lgs}_1(\valpha_p)\phi^{lgs}_1(\valpha_q)> \\  <\phi^{lgs}_2(\valpha_p)\phi^{lgs}_2(\valpha_q)> \\  <\phi^{lgs}_1(\valpha_p)\phi^{lgs}_2(\valpha_q)>\end{array}\right\} &=& \frac{7.791}{ \int_0^{\infty} C_n^2(h)\mathrm{d}h}\left(\frac{D}{r_0}\right)^{\frac{5}{3}} \nonumber \\
&\times& \int_0^L \mathrm{d}h~[\zeta(h)]^{\frac{5}{3}.}C_n^2(h)~[\left\{\begin{array}{r}1 \\ 1 \\ 0\end{array}\right\}HJ2(14/3,2,0,1, \rho^{pq}(h)) \nonumber \\ 
&& - \left\{\begin{array}{r} \cos(2\theta^{pq}) \\ -\cos(2\theta^{pq}) \\ \sin(2\theta^{pq})\end{array}\right\}  HJ2(14/3,2,2,1, \rho^{pq}(h))] 
\end{eqnarray}
\begin{eqnarray}
\left\{\begin{array}{c}  <s_k^{x}(\valpha_p)\phi^{lgs}_1(\valpha_q)> \\  <s_k^{y}(\valpha_p)\phi^{lgs}_2(\valpha_q)> \\ <s_k^{x}(\valpha_p)\phi^{lgs}_2(\valpha_q)> \\ <s_k^{y}(\valpha_p)\phi^{lgs}_1(\valpha_q)> \end{array}\right\} &=& \frac{0.620}{ \int_0^{\infty} C_n^2(h)\mathrm{d}h}\left(\frac{D}{r_0}\right)^{\frac{5}{3}}\left(\frac{\lambda}{R_s}\right) \nonumber \\
&& \hskip-40pt \times \int_0^L \mathrm{d}h~[\zeta(h)]^{\frac{5}{3}.}C_n^2(h)~[\left\{\begin{array}{r}1 \\ 1 \\ 0 \\ 0 \end{array}\right\}HJ3(11/3,1,2,0,R_s/R,1,\rho_k^{pq}(h)) \nonumber \\ 
&& - \left\{\begin{array}{r} \cos(2\theta_k^{pq}) \\ -\cos(2\theta_k^{pq}) \\ \sin(2\theta_k^{pq}) \\ \sin(2\theta_k^{pq})\end{array}\right\}HJ3(11/3,1,2,2,R_s/R,1,\rho_k^{pq}(h))]  
\end{eqnarray}
where $\displaystyle [\rho_{kl}, \theta_{kl}]=\ro_l-\ro_k$, $\displaystyle [\rho^{pq}(h), \theta^{pq}] = \frac{h}{R\zeta(h)}(\valpha_q - \valpha_l)$,  $\displaystyle [\rho^{pq}_{k}(h),\theta_{k}^{pq}(h)] = \frac{h}{R\zeta(h)}(\valpha_q - \valpha_p) - \ro_k$ and $\displaystyle [\rho^{pq}_{kl}(h),\theta_{kl}^{pq}(h)] = \frac{h}{R\zeta(h)}(\valpha_q - \valpha_p) + \ro_l-\ro_k$\\
Similarly $\mathrm{Cov}(\vphi,\vs)$ represents the cross-correlation between the slopes and the tip/tilt removed science star turbulent phase. The elements $C_{s\phi}^{x}$, $C_{s\phi}^{y}$ of the matrix are defined by:
\begin{eqnarray}
C_{s\phi}^{x}&=&<s_k^x(\valpha_p)\phi_j>-\left(\frac{\lambda}{\pi{R}}\right)<\phi^{lgs}_1(\valpha_p)\phi_j>\\
C_{s\phi}^{y}&=&<s_k^y(\valpha_p)\phi_j>-\left(\frac{\lambda}{\pi{R}}\right)<\phi^{lgs}_2(\valpha_p)\phi_j>
\end{eqnarray}
The computation of these moments are provided in Appendix B. In the specific case of SH circular subapertures, their expression involves $H3J$ function: 
\begin{eqnarray}
&&\left\{\begin{array}{c} <s_k^{x}(\valpha_p)\phi_j> \\  <s_k^{y}(\valpha_p)\phi_j> \end{array}\right\} = s_{n,m}\frac{0.310}{\int_0^{\infty} C_n^2(h)\mathrm{d}h}\left(\frac{D}{r_0}\right)^{\frac{5}{3}} \left(\frac{\lambda}{R_s}\right) \\
&& \hskip10pt \times \int_0^L \mathrm{d}h~C_n^2(h)~[\left\{\begin{array}{r} \beta^{x}_{|m|-1}(\theta_k^p(h)) \\ \beta^{y}_{|m|-1}(\theta_k^p(h)) \end{array}\right\}HJ3(11/3,1,n+1,|m|-1,\zeta(h)R_s/R,1,\rho^p_k(h)) \nonumber \\ 
&& \hskip70pt + \left\{\begin{array}{r} -\beta^{x}_{|m|+1}(\theta_k^p(h)) \\ \beta^{y}_{|m|+1}(\theta_k^p(h))\end{array}\right\}HJ3(11/3,1,n+1,|m|+1,\zeta(h)R_s/R,1,\rho^p_k(h)] \nonumber
\end{eqnarray}
\begin{eqnarray}
&&\left\{\begin{array}{c} <\phi_1^{lgs}(\valpha_p)\phi_j> \\  <\phi_2^{lgs}(\valpha_p)\phi_j> \end{array}\right\} = s_{n,m}\frac{3.986}{\int_0^{\infty} C_n^2(h)\mathrm{d}h}\left(\frac{D}{r_0}\right)^{\frac{5}{3}} \\
&& \hskip10pt \times \int_0^L \mathrm{d}h~[\zeta(h)]^{-1}C_n^2(h)~[\left\{\begin{array}{r} \beta^{x}_{|m|-1}(\theta^p) \\ \beta^{y}_{|m|-1}(\theta^p) \end{array}\right\}HJ3(14/3,2,n+1,|m|-1,\zeta(h),1,\rho^p(h)) \nonumber \\ 
&& \hskip100pt + \left\{\begin{array}{r} -\beta^{x}_{|m|+1}(\theta^p) \\ \beta^{y}_{|m|+1}(\theta^p)\end{array}\right\}HJ3(14/3,2,n+1,|m|+1,\zeta(h),1,\rho^p(h)] \nonumber
\end{eqnarray}
where $\displaystyle [\rho^{p}_{k}(h),\theta_{k}^{p}(h)] = \frac{h\valpha_q}{R} + \zeta(h)\ro_k$ and $\displaystyle [\rho^{p}(h),\theta^{p}] = \frac{h\valpha_q}{R}$. \\
Finally, $\mathrm{Cov}(\vepsilon)$ represents the additive noise covariance. Assuming identical noises for all SH  and that the noises are independent between two different subapertures, the covariance matrix can be rewritten $\mathrm{Cov}(\vepsilon) = \sigma^2_\epsilon.\mathrm{Id}$, where $\mathrm{Id}$ is the identity matrix and $\sigma^2_\epsilon$ is the quadratic sum of the photon ($\sigma_p^2$) and detector noises ($\sigma_d^2$). Rousset \cite{rousset_1} has given an expression for both noises, in the case of SH wavefront sensors:
\begin{eqnarray}
\sigma_p^2 &=& \left(\frac{\pi}{\sqrt{2}}\right)^2 \frac{1}{N_{ph}} \left(\frac{X_T}{X_D}\right)^2\\
\sigma_d^2 &=& \left(\frac{\pi}{\sqrt{3}}\right)^2 \frac{\sigma^2_{e^-}}{N^2_{ph}} \left(\frac{4X_T^2}{X_D}\right)^2
\end{eqnarray}
where $N_{ph}$ is the number of photons per subaperture, $\sigma_{e^-}$ is the detector noise rms per pixel, and $X_T$, $X_D$ are the full width half maximum (in pixels) of respectively the turbulent and diffraction-limited subaperture image spots. As it is beyond the scope of this paper, the previous equations do not take into account the effect of the laser spot  elongation on the SH subapertures due to the parallax effect and the non-zero thickness of the layer where the spot is created. This additional effects that varies with the radial location of the LGS can  be taken into consideration by replacing previous equations with that of e.g. Bechet \textit{et al.} \cite{bechet_1} (see Eq. (6) of their paper).

\subsection{Discrete turbulent layers: matrix approach} \label{sec_mat_approach}
\begin{figure}[htbp]
\begin{center}\includegraphics[width=0.45\textwidth]{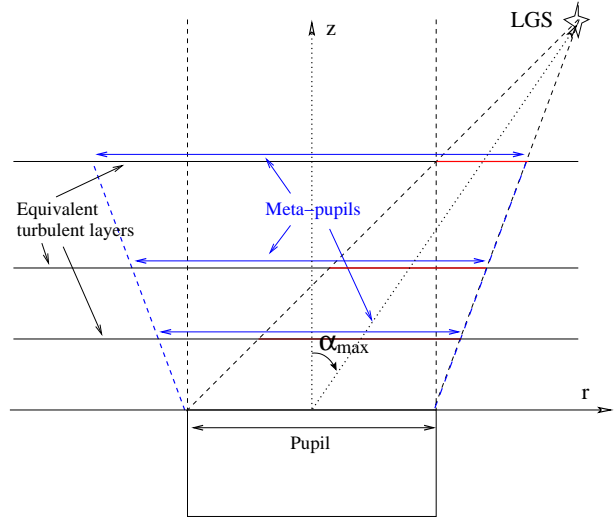}
\caption{\label{fig_sketch2} Same as Fig. (\ref{fig_sketch1}) in the case of discrete turbulent layers. The location of the LGS defines for each layer a so-called \textit{metapupil} of radius the maximum between $R$ and $R+h(|\valpha_{max}|-\frac{R}{L})$.}
\end{center}
\end{figure}
We now assume that the turbulent medium can be modelled by a discrete sum of $N_{el}$ equivalent, statistically independent turbulent layers of thickness $\Delta{h}$, as sketched in Fig. (\ref{fig_sketch2}). In such a case, Eqs. (\ref{eq_zernphi}) and (\ref{eq_zernphilgs}) can be respectively rewritten as:
\begin{equation}
\Phi(R\ro) = \sum_{j=4}^{\infty}\sum_{k=1}^{N_{el}}\phi_j(h_k) Z_j(\ro)\label{eq_zernphi_h}
\end{equation}
\begin{equation}
\Phi^{lgs}(R\ro, \valpha_p) = \sum_{j=1}^{\infty}\sum_{k=1}^{N_{el}}\phi^{lgs}_j(\valpha_p, h_k) Z_j(\ro)\label{eq_zernphilgs_h}
\end{equation}
where $\vphi(h_k)$, $\vphi^{lgs}(\valpha_p, h_k)$ are the Zernike coefficients for respectively the science and LGS phase of the $k^{th}$ turbulent layer.  For each layer, the outer part of all the LGS cone beams together defines the limits of a so-called \textit{meta-pupil} \cite{ragazzoni_1} which covers the turbulence crossed by both the science and LGS wavefronts at that layer. If $\valpha_{max}$ is the largest angular location of the LGS network, the size of the meta-pupil $R^{\mathcal{M}}(h)$ is defined as:
\begin{equation}
R^{\mathcal{M}}(h) = \left\{\begin{array}{cl} R & \mathrm{if}~|\valpha_{max}| \le \frac{R}{L} \\ R + h\left(|\valpha_{max}|- \frac{R}{L}\right) & \mathrm{if}~|\valpha_{max}| > \frac{R}{L} \end{array} \right.
\end{equation}
We call $\Phi^{\mathcal{M}}$, $\vphi^{\mathcal{M}}(h_k)$ the phase and its associated Zernike coefficients defined over the metapupils of each turbulent layer. Ragazzoni \textit{et al.} \cite{ragazzoni_1} have shown that there exists linear procedures (i.e. matrices) that allow one to deduce the Zernike coefficients of the science and LGS wavefronts from those of the meta-pupils. We call these matrices $\mathcal{W}_{h_k}$ and $\mathcal{L}^{\alpha_p}_{h_k}$ respectively. They satisfy: 
\begin{eqnarray}
\vphi(h_k) &=& \mathcal{W}_{h_k}.\vphi^{\mathcal{M}}(h_k) \label{eq_Wh}\\
\vphi^{lgs}(\valpha_p, h_k) &=& \mathcal{L}^{\alpha_p}_{h_k}.\vphi^{\mathcal{M}}(h_k) \label{eq_Lh}
\end{eqnarray}
Several techniques are available in the literature to evaluate the coefficients of both matrices. As $\mathcal{W}$ deals with pupil scaling (from $R^{\mathcal{M}}$ to $R$), one can indifferently use the methods of \cite{schwiegerling_1, campbell_1, shu_1}.  The calculation of $L$ is  more complex since it requires pupil translation (from $0$ to $h|\valpha_{max}|$) in addition to pupil scaling (from $R^{\mathcal{M}}$ to $R[1-h/L]$). Different, however equally working ways of solving the problem are available \cite{bara_1, lundstrom_1, tatulli_1}.
Using altogether Eqs. \ref{eq_sh}, \ref{eq_zernphi_h}, \ref{eq_zernphilgs_h}, \ref{eq_Wh}, \ref{eq_Lh}, we get :
\begin{eqnarray}
\vphi &=& \sum^{N_{el}}_{k=1}\mathcal{W}_{h_k}.\vphi^{\mathcal{M}}(h_k)\\
\vs &=& D_{\infty}\sum^{N_{el}}_{k=1}\mathcal{L}^{\alpha}_{h_k}.\vphi^{\mathcal{M}}(h_k) 
\end{eqnarray}
where $\mathcal{L}^{\alpha}$ is the block-diagonal matrix including all matrices $\mathcal{L}^{\alpha_p}$, $p \in [1..N_{lgs}]$.\\
As a result, in the case of an atmospheric model made of statistically independent discrete turbulent layers, the covariance matrices required to compute the residual error of Eq. (\ref{eq_sig2res}) are given by:
\begin{eqnarray}
\mathrm{Cov}(\vphi) &=& \sum^{N_{el}}_{k=1} \mathcal{W}_{h_k} \mathrm{Cov}(\vphi^{\mathcal{M}}_{h_k}) \mathcal{W}_{h_k}^T \label{eq_covphi}\\
\mathrm{Cov}(\vs) &=&  D_{\infty} \sum^{N_{el}}_{k=1}\mathcal{L}^{\alpha}_{h_k}\mathrm{Cov}(\vphi^{\mathcal{M}}_{h_k}) [\mathcal{L}^{\alpha}_{h_k}]^TD_{\infty}^T \label{eq_covs}\\
\mathrm{Cov}(\vphi,\vs)  &=& D_{\infty}\sum^{N_{el}}_{k=1}\mathcal{L}^{\alpha}_{h_k}\mathrm{Cov}(\vphi^{\mathcal{M}}_{h_k})\mathcal{W}_{h_k}^T = \sum^{N_{el}}_{k=1} \mathcal{W}_{h_k} \mathrm{Cov}(\vphi^{\mathcal{M}}_{h_k})[\mathcal{L}^{\alpha}_{h_k}]^TD_{\infty}^T \label{eq_covphis}
\end{eqnarray}
where $\mathrm{Cov}(\vphi^{\mathcal{M}}_{h_k})=<\vphi^{\mathcal{M}}(h_k)[\vphi^{\mathcal{M}}(h_k)]^T>$ is the covariance matrix of the turbulent phase of the $k^{th}$ atmospheric layer for which an expression is given by Noll \cite{noll_1} with $D=2R^{\mathcal{M}}(h)$.

\subsection{Minimum Mean Square Error: optimal control matrix}\label{subsec_mmse}
Assuming Gaussian statistics for the noise, the classical estimator  of the reconstructed phase in the least square sense is defined by the generalized inverse of $D_{\infty}$:
\begin{equation}
M_{svd}=\left[D_{\infty}^T.D_{\infty}\right]^{-1}.D_{\infty}^T \label{eq_Msvd}
\end{equation}
In theory, the number of columns $N_z$ in $D_{\infty}$ is infinite as is the number of polynomials in the Zernike basis. In practice, if we set $N_z$ to a high number (i.e. $N_z \gg M_{s}$) the matrix $[D_{\infty}^T.D_{\infty}]$ becomes ill-conditioned because of frequency aliasing due to the finite size of the subapertures and its inversion introduces an unacceptable increase in the noise. If on the contrary, we compute $D_{\infty}$ with a low number of Zernike modes (typically $N_z \simeq M_{s}/2$), we introduce a \textit{modelling} error \cite{veran_1} as the description of the phase on the Zernike basis is incomplete. The value of $N_z$ (which moreover depends on the SNR of the measurements) must be chosen carefully in order to obtain a fair trade-off between both aliasing and modelling errors and in practice this method reveals itself unsatisfactory. \\
To circumvent this problem Fusco \cite{fusco_1} has proposed to compute the control matrix by minimizing the residual variance  by exploiting the prior knowledge of both the statistics of the phase ($\mathrm{Cov}(\vphi)$) and the noise of the slopes measurements ($\mathrm{Cov}(\vepsilon)$). The so-called Minimum Mean Square Error (MMSE) estimator is then derived such that $\mathrm{d}\sigma^2_{res}/\mathrm{d}M=0$. It turns out that \cite{fusco_1}:
\begin{equation}
M_{ngs}=\mathrm{Cov}(\vphi)D^T_{\infty}\left[D_{\infty}\mathrm{Cov}(\vphi)D^T_{\infty} + \mathrm{Cov}(\vepsilon)\right]^{-1} \label{eq_Mao}
\end{equation}
The previous equation is however derived from the hypothesis of a single, natural (e.g. plane wavefront), 
 on-axis guide star, hence the above expression is optimal only for this particular case. For providing a generalization of the MMSE estimator for single/multiple spherical wavefront guide stars located at any angular positions in the sky, we simply perform the matrix derivation with respect to $M$ of the residual variance of Eq. (\ref{eq_sig2res}), introducing the definition of Eqs. (\ref{eq_covphi}, \ref{eq_covs}, \ref{eq_covphis}) for the respective covariances. We finally obtain:  
\begin{equation}
M_{lgs} = \sum^{N_{el}}_{k=1} \mathcal{W}_{h_k} \mathrm{Cov}(\vphi^{\mathcal{M}}_{h_k})[\mathcal{L}^{\alpha}_{h_k}]^TD_{\infty}^T \left[D_{\infty}\sum^{N_{el}}_{k=1}\mathcal{L}^{\alpha}_{h_k}\mathrm{Cov}(\vphi^{\mathcal{M}}_{h_k}) [\mathcal{L}^{\alpha}_{h_k}]^TD_{\infty}^T + \mathrm{Cov}(\vepsilon)\right]^{-1} \label{eq_controlmatrix}
\end{equation}
This equation is valid for any number of guide stars. In the case of multiple guide stars, we find here the same expression as for the Multi-Conjugate Adaptive Optics (MCAO) control matrix \cite{fusco_2} but altered for the specific case of LTAO sub-class, which works with a projection over a single deformable mirror and for a single direction of interest at the center of the field \cite{neichel_1}.\\
We precise that previous optimal control matrices refer to the minimization of the residual phase error for \textit{open-loop} AO correction, as it is usually investigated in the litterature. For closed-loop operations, one needs to take into account the feedback towards the DM that drives to null the signal generated by the guide star wavefront sensing. In other words, the error that is contemplated to be minimized in closed-loop is defined by $<\Delta_s^2>$ with \cite{wallner_1}:
\begin{equation}
<\Delta_s^2> = <(\widehat{\vs}-D_{\infty}\widetilde{\vphi})^2> =  <(\widehat{\vs}-D_{\infty}M[\widehat{\vs} + \vepsilon])^2>
\end{equation}
In that case, the MMSE closed-loop control matrix $M^{cl}_{lgs}$ takes a slighty different form:
\begin{equation}
M^{cl}_{lgs} = \sum^{N_{el}}_{k=1}\mathcal{L}^{\alpha}_{h_k}\mathrm{Cov}(\vphi^{\mathcal{M}}_{h_k}) [\mathcal{L}^{\alpha}_{h_k}]^TD_{\infty}^T\left[D_{\infty}\sum^{N_{el}}_{k=1}\mathcal{L}^{\alpha}_{h_k}\mathrm{Cov}(\vphi^{\mathcal{M}}_{h_k}) [\mathcal{L}^{\alpha}_{h_k}]^TD_{\infty}^T + \mathrm{Cov}(\vepsilon)\right]^{-1} \label{eq_controlmatrix_cl}
\end{equation}
 In the following we will focuses on the open-loop definitions. This choice does however not affect the conclusions of our analysis.

\section{The single LGS case}\label{sec_onelgs}
We present in this section the theoretical performance of AO correction using one LGS and the optimal MMSE wavefront reconstruction of Eq. (\ref{eq_controlmatrix}). They are analysed alternatively in terms of phase residual error ($\sigma_{res}$) given in radians, or Strehl ratio ($SR(\lambda_{im})$) for a given imaging wavelength $\lambda_{im}$ with  $SR(\lambda_{im}) = \exp[-\sigma_{res}^2(\lambda/\lambda_{im})^2]$. In the following examples, we have enforced a SNR of $100$ (the SNR being defined as the ratio between the turbulence variance and the noise variance \cite{fusco_2}) such that we focus here on the performance limitations due to the combined effects of partial AO correction provided by the WFS (the fitting error) and focal anisoplanatism (cone effect). The global evolution of the error with the number of photoevents is presented in Sect. (\ref{sec_compare}).  \\
In the first graph of Fig. (\ref{fig_1lgs}), we also compare the results obtained with the general approach of Sect. (\ref{sec_gen_approach}) (dashed lines) with that of the matrix study of Sect (\ref{sec_mat_approach}) (solid lines). Although in good agreement, we  note a small ($\sim 10^{-2}$rad for $D/r_0=1$) however systematic difference between the two methods. This is explained by the fact that the matrix approach, in order to estimate the covariance matrices $\mathrm{Cov}(\vphi)$, $\mathrm{Cov}(\vs)$ and $\mathrm{Cov}(\vphi,\vs)$, requires the effective computation of the interaction matrix $D_{\infty}$ which in practice will only take into account a finite number of Zernike modes (that we have set in this paper to $N_{z}=406$, that is up to $n_z=27$ radial degrees). On the contrary, the general technique fully computes the same covariance matrices without going through the Zernike basis description, i.e. without modelling error. An upper limit of the discrepancy between both methods can thus be roughly estimated from the remaining turbulent error of the modes not considered in the matrix approach. The turbulent variance of uncorrected Zernike polynomials from $n_z + 1$ to $\infty$ is given by Conan \cite{conan_2}  and writes $\Delta^2_{n_z} \simeq 0.458 (n_z + 1)^{-5/3}(D/r_0)^{5/3}$. For $n_z=27$ and $D/r_0=1$, it comes $\Delta_{n_z} \simeq 4\times10^{-2}$ rad which is consistent with our results.\\
In order to validate our calculations, we have also built a quick simulation tool that models the problem by (i) generating a sample of $N_s$ random screen phases (here $Ns = 500$) following Kolmogorov turbulence (using Roddier method \cite{roddier_n_1}) and (ii) numerically computing the slopes of the (tip-tilt removed) phases over each subaperture. The simulated residual error together with its statistical dispersion are overplotted (symbols and error bars) and match the theoretical curves, hence confirming our analytical approaches.\\
In the following, the parameters are fixed as $D=2$m, $r0=12$cm (in R band), $M_s=69$, $h=[0.01, 5, 12]$km, $\Delta{h}=0.5$km (the layer thickness being required to compute $r_0$) and $\alpha=0.5D/L$, unless when taken as variables or mentioned otherwise. 

\subsection{Focal anisoplanatism and fitting error}
\begin{figure*}[htbp]
\begin{center}\begin{tabular}{c}
\includegraphics[width=0.9\textwidth]{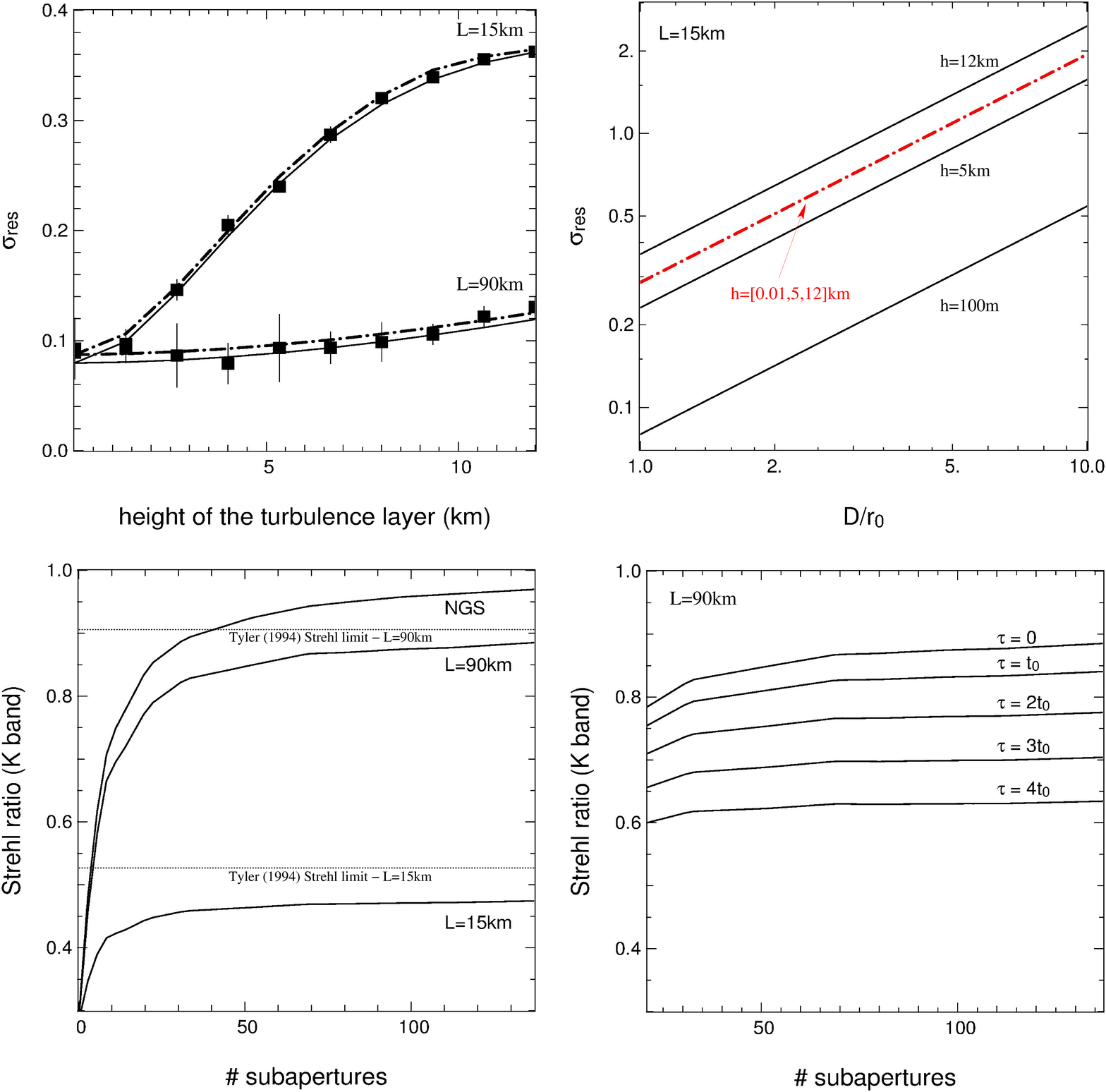}
\end{tabular}
\caption{\label{fig_1lgs} Top left: residual error as a function of the turbulent layer altitude for both $L=90$km and $L=15$km, with $D/r_0=1$. The methods used to compute the error are the matrix approach (solid line), the general approach (dashdotted line) and through simulations (symbols + statistical dispersion). Top right: residual error as a function of the turbulence strength, for various (single layers and combination of 3 layers) atmospheric profiles. Bottom: K-band Strehl ratio as a function of the number of subapertures, for different GS altitude (left) and AO control loop time delays (right).}
\end{center}
\end{figure*}
\textit{Turbulent layer vs. LGS altitude:} Figure (\ref{fig_1lgs}, top left) shows the behavior of the residual error as a function of the height of the turbulent layer (one layer considered here), assuming $D/r_0=1$. As the altitude of the turbulent layer increases, the fraction of the turbulence crossed by the LGS beam decreases, hence the quality of the correction. This graph illustrates the well-known cone effect due to the finite altitude of the LGS star. When the LGS is high in the sky, that is significantly higher than the upper turbulent layer, the cone effect remains fairly small. On the contrary when the LGS lies close to the turbulence, the performance can be degraded up to a factor of $\sim 3$ between layers at $1$km and $12$km. When the turbulent layer is above the LGS, no correction is performed and the error saturates at the Noll value of Eq. (\ref{eq_noll}), that is $\sigma_{res} \simeq 0.37$ rad for $D/r_0=1$. Figure (\ref{fig_1lgs}, top right) displays the residual error as a function of ${D}/{r_0}$ for different turbulent layers, respectively $h=0.01$km, 5km, 12km, and a combination of these three with $C_n^2$ strengths chosen such that $r_0$ keeps the same value as that of single layer profiles.For obtaining $r_0=12$cm in R band, We thus have set $C_n^2=1.7 \times 10^{-15}\mathrm{m}^{-2/3}$ for one layer, and equal $C_n^2$ values of $5.7 \times 10^{-16}\mathrm{m}^{-2/3}$ for three layers. We can see that, like the residual error in classical AO correction, the LGS AO is  following a $({D}/{r_0})^{-5/6}$ law. In the case of the 3-layer turbulent profile (dashdotted line), the error is mostly driven by the higher layer of the turbulence where the cone effect is the strongest. \\ 
\textit{Cone effect vs. fitting error:} Figure (\ref{fig_1lgs}, bottom left) shows the evolution of the K-band Strehl ratio with the number of subapertures (and consequently the size of the subapertures). As expected the Strehl increases with $M_s$, since the subaperture diameter decreases and the WFS provides a tighter sampling of the incoming wavefront. However, because  most of the turbulent energy is contained in the low modes, the Strehl slowly bend towards a flatter curve and the gain in performance becomes progressively marginal. For guide stars located at finite altitude, the AO correction is in addition severely limited by the cone effect that causes an overall loss in the performance, roughly of $10\%$ ($L=90$km) and $50\%$ ($L=15$km) of the K-band Strehl ratio expected for a natural guide star. Focal anisoplanatism also drives to a stronger saturation of the performance so that it becomes worthless to increase the number of subaperture at some point. This is especially true for low altitude LGS where the SR reaches a plateau for a  small number of subapertures ($M_s \simeq 20-30$ for $D=2$m).Tyler \cite{tyler_2} has provided the residual phase variance due to focal anisoplanatism for the case of perfect (i.e perfect wavefront sensor) on-axis correction using LGS. He has shown that this variance could be written under the form $(D/d_0)^{5/3}$ where $d_0$ is the so-called effective diameter of a LGS compensated imaging system, and is given by Eq. (61) of his paper. This result can be translated into Strehl ratio upper limits, that is the maximum achievable performance using LGS AO correction. In our case, we find  $SR=0.91$ and $SR=0.53$ in K-band, for respectively LGS at $90$km and $15$km, which is consistent with our results presented in Figure (\ref{fig_1lgs}, bottom left).\\
\textit{Control loop delay:} We  investigate here the effect of a time delay $\tau$ in the AO loop.  In the case of our 3-layers profile, we obtain an average wind speed of $\overline{v} \simeq 18.5$m/s and a (R band) coherence time of $t_0 \simeq 2$ ms. Fig. (\ref{fig_1lgs}, bottom right) shows how  the performance is degrading with an increasing time delay. Such an effect is expected since time delay translates into spatial decorrelation between the science and guide star, hence damaging the AO correction. As a consequence the SR continuously degrades as the time delay increases and performance can undergo severe loss in cases of integration times significantly higher than the coherence time of the atmosphere, with e.g. a K-band SR loss of $\sim 25\%$ for $\tau = 4t_0$. These results are in agreement with Min \& Yi study\cite{min_1}.

\subsection{Analysis of the optimal reconstruction} 
\begin{figure*}[htbp]
\begin{center}\begin{tabular}{c}
\includegraphics[width=0.9\textwidth]{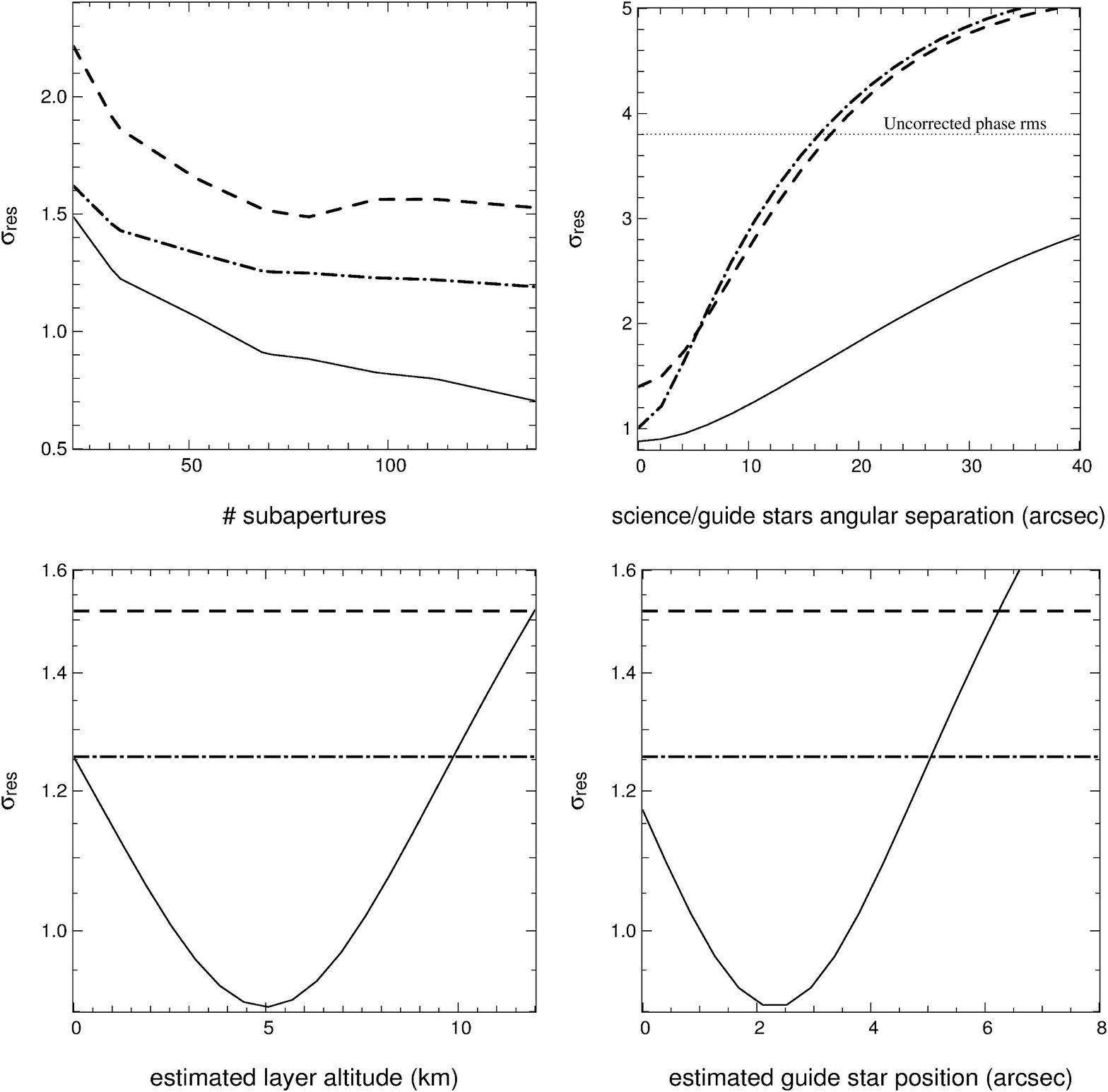}
\end{tabular}
\caption{\label{fig_1lgs_suite} Top: residual error as a function of the number of subapertures (left) and the angular separation between the science and guide star (right). Bottom:  residual error as a function of the \textit{estimated} turbulence layer altitude (left) and LGS angular position (right). The correct values have been set to $h=5$km, $\alpha=0.5D/L$. Different control matrices are considered: SVD method (Eq. (\ref{eq_Msvd}), dashed line), MMSE NGS (Eq. (\ref{eq_Mao}), dashdotted line) an MMSE LGS (Eq. (\ref{eq_controlmatrix}), solid line).}
\end{center}
\end{figure*}
Figure (\ref{fig_1lgs_suite}, top-left) compares the levels of correction with respect to the chosen control matrix. We can see an improvement as we go from the SVD matrix (Eq. (\ref{eq_Msvd})) to MMSE methods, both for on-axis NGS (Eq. (\ref{eq_Mao})) as well as LGS (Eq. (\ref{eq_controlmatrix})) matrices. The latter case provides the best results with an improvement of a factor of $\sim 1.5$ with respect to the SVD reconstruction method. For the SVD reconstruction, we have set the number of Zernike polynomials to be equal to half the number of subapertures. When the latter number is increased (and hence the number of Zernikes), we can see that at some point the performance of the SVD estimator starts to decrease, the residual error climbing up again. This behavior illustrates the inability of the WFS to ``see'' some particular modes when no regularization has been performed. Furthermore, as shown in Fig. \ref{fig_1lgs_suite} (top-right), the $M_{ltao}$ control matrix allows to significantly improve the effective isoplanatic patch of the LGS star (defined as the range of angular distance over which the residual error is lower than that of the uncorrected one) by at least a factor of $3$ as $\alpha>40^{''}$ when  the MMSE LGS method is used in place of $\alpha \simeq15^{''}$ for the SVD/MMSE NGS reconstruction techniques.\\
Using the MMSE reconstruction in order to optimize LGS AO correction, however requires an \textit{a priori} knowledge of both the altitude of the turbulent layers and the location of the guide star; parameters that are difficult to estimate precisely and can also slowly vary with time. Figure \ref{fig_1lgs_suite} (bottom) investigates (for the one-layer atmospheric model) the robustness of the technique to an error in the estimation of the layer altitude and the angular location of the LGS. It shows that MMSE LGS estimator can tolerate large uncertainties of $\Delta{h} \sim 5$km and $\Delta\alpha \sim 3^{''}$ before reaching similar performance to that of $M_{ngs}$ control matrix. The range even widens when compared to the standard SVD technique. These ranges depend neither on the true value of the turbulent layer height nor that of the LGS angular location since the MMSE LGS curves will shift only along the x-axis as a function of these values.  

\section{Tomography}\label{sec_tomo}
\subsection{Compensating the cone effect with a network of guide stars}
\begin{figure*}[htbp]
\begin{center}\begin{tabular}{c}
\includegraphics[width=0.9\textwidth]{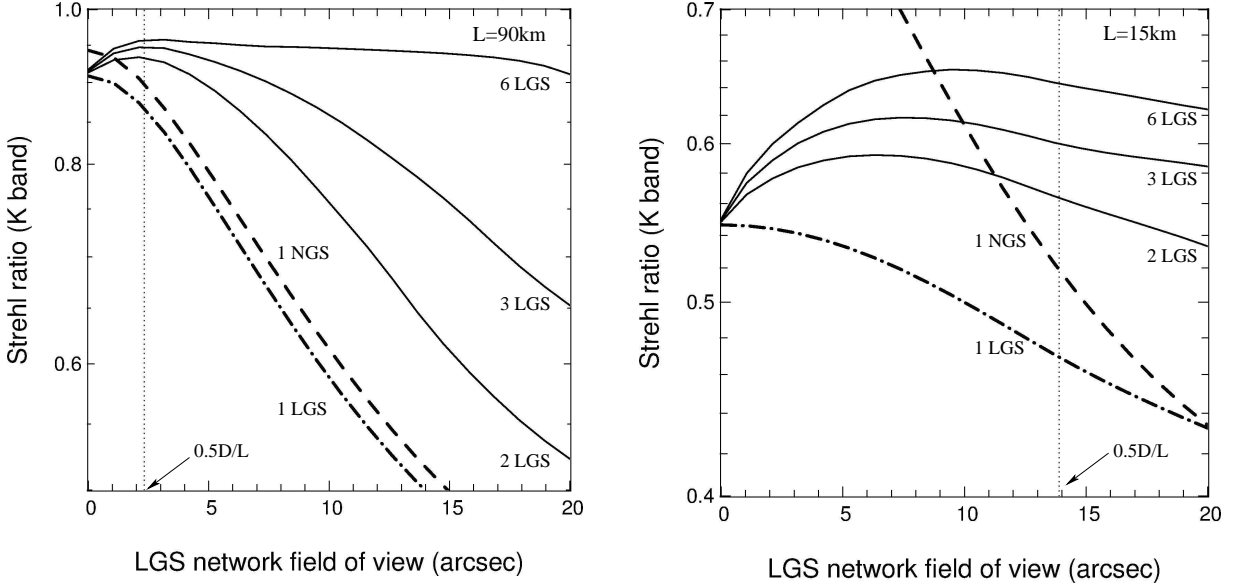}
\end{tabular}
\caption{\label{fig_ltao} K-band Strehl ratio as a function of LGS circle radius, for different numbers of LGS, in both cases of high altitude (L=$90$km, left) and low altitude (L=$15$km, right) guide stars. The dashdotted line displays the corresponding single LGS case whereas the dashed line illustrates the single NGS case. The dotted vertical line shows the LGS angle corresponding to the edge of the telescope. Observations with $2$, $3$ and $6$ LGS are considered, as indicated on the plots.}
\end{center}
\end{figure*}
To circumvent the cone effect limitation, one can use a network of LGSs located at different positions in the sky and carrying out a tomographic reconstruction of the atmosphere. In the following, the LGSs will be radially distributed on a circle the radius of which (so-called LGS field of view) can vary. Figure (\ref{fig_ltao}) shows the K-band SR as a function of the LGS field of view, for increasing number of guide stars, for both cases of Sodium ($L=90$km) and Rayleigh ($L=15$km) lasers. As expected, using several LGSs instead of one allows an increase in the quality of the correction. And the improvement is all the more significant when the single LGS is launched off-axis. For the Sodium laser case, we can see that $4$ LGSs are enough to fully cancel out the cone effect and reach the performance of an on-axis natural guide star. On the contrary, the cone effect can only be partially compensated when using Rayleigh lasers, $6$ LGSs allowing to reach $\sim 70\%$ of the K-band Strehl ratio of the on-axis natural guide star.\\
For high altitude LGS system, the optimal LGS FOV is strongly marked and the performance can be severely degraded when the LGS circle deviate from this specific radius, especially when a few number of LGSs are used. The optimum is found to be for $\alpha \sim 0.5D/L$, that is when the circle which the LGSs draw on sky matches with the edge of the telescope aperture. This empirical law can also be deduced from rough geometrical considerations  \cite{lelouarn_1} noticing that  $\alpha = 0.5D/L$ is the minimum angle that enable to encompass the full volume of turbulence (the outer part of the LGS beams in that case being superimposed to that of the science star beam). Tokovinin \textit{et al.} \cite{tokovinin_1} have also found the same optimum from their numerical code (see for e.g. in Fig. (4) of their paper with an optimal radius of $\sim 9^{''}$ for $d=8$m, $L=90$km, in the case of 3LGS). The SR optimum is however not as sharp when the number of LGS is  bigger than the number of turbulent layers. Also, the rule is valid only when the LGS are significantly higher than the upper atmospheric layer, roughly when $h_{upper} \le L/2$. For low altitude LGS such as Rayleigh stars, the situation is less clear. As the cone effect is stronger, the optimal angular radius will depend on the altitude of the upper turbulent layer. In that case, a theoretical analysis of the performance is suitable for \textit{a priori} estimating the best radius of the LGS network  according to the atmospheric properties of the observational site (most of all the altitude of the upper layer). 

\subsection{Validity of atmospheric equivalent layers modelling}\label{sec_el}
\begin{figure*}[htbp]
\begin{center}\begin{tabular}{c}
\includegraphics[width=0.95\textwidth]{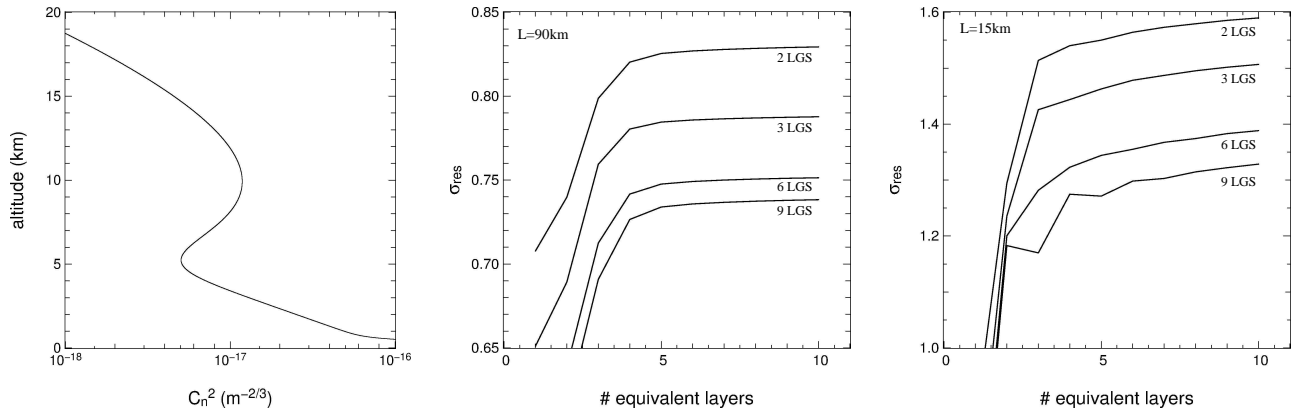}
\end{tabular}
\caption{\label{fig_ltao2}Left: input Hufnagel $C_n^2(h)$ profile. Middle: residual error as a function of the number of equivalent layers, for different numbers of laser spots at $L=90$km, as indicated on the curves. Right: same as previously, but for $L=15$km.}
\end{center}
\end{figure*}
The use of LTAO reconstruction requires the turbulent profile to be decomposed in discrete thin layers (so-called equivalent layers \cite{fusco_2}), in order to achieve atmosphere tomography. We analyse in this section the validity of such a decomposition and estimate how  many layers are needed to correctly describe the effects of a given continuous turbulent profile. We consider a classical Hufnagel continuous (night) profile \cite{parenti_1}, as shown in Fig. (\ref{fig_ltao2}, left), adjusting parameters in order to obtain $r_0=12$cm in R band. We then slice this profile in $N_{el}$ equally thick zones and compute for each zone the height of the equivalent layer, such that the $k^{th}$ altitude verifies:
\begin{equation}
h_k = \frac{\int_{h_{min}(k)}^{h_{max}(k)} h C_n^2(h) \mathrm{d}{h}}{\int_{h_{min}(k)}^{h_{max}(k)} C_n^2(h) \mathrm{d}{h}}
\end{equation}
where $h_{min}(k)$ and $h_{max}(k)$ are the lower and upper limits of the $k^{th}$ $C_n^2$ zone. The turbulence associated to this layer is:
\begin{equation}
C_n^2(h_k)  = \frac{\int_{h_{min}(k)}^{h_{max}(k)} C_n^2(h) \mathrm{d}{h}}{\Delta{h}}
\end{equation}
Fig. (\ref{fig_ltao2}, middle and right) shows the evolution of the error as a function of the chosen number of equivalent layers $N_{el}$. It is clear that, independent of the number and altitudes of the LGSs, the residual error quickly reaches a plateau, showing that only a few layers ($\sim 4$) are sufficient for a proper modelling of the LTAO correction. The plateau is less pronounced in the case of low altitude LGS, although the relative error on the phase residual estimate remains  $\lsim 5\%$ when modelling the atmosphere with 4 EL instead of 10. This translates into a relative error of $\lsim 2\%$ in the estimation of the associated K-band Strehl ratio. Our study thus theoretically validates the relevance of the equivalent layer approach. It is consistent with Fusco analysis who concluded that \textit{``only a small number of layers are needed to obtain a good precision on the statistical behavior of the turbulent phase''} \cite{fusco_1}.

\section{Sodium vs. Rayleigh guide stars}\label{sec_compare}
\begin{figure*}[htbp]
\begin{center}\begin{tabular}{c}
\includegraphics[width=0.9\textwidth]{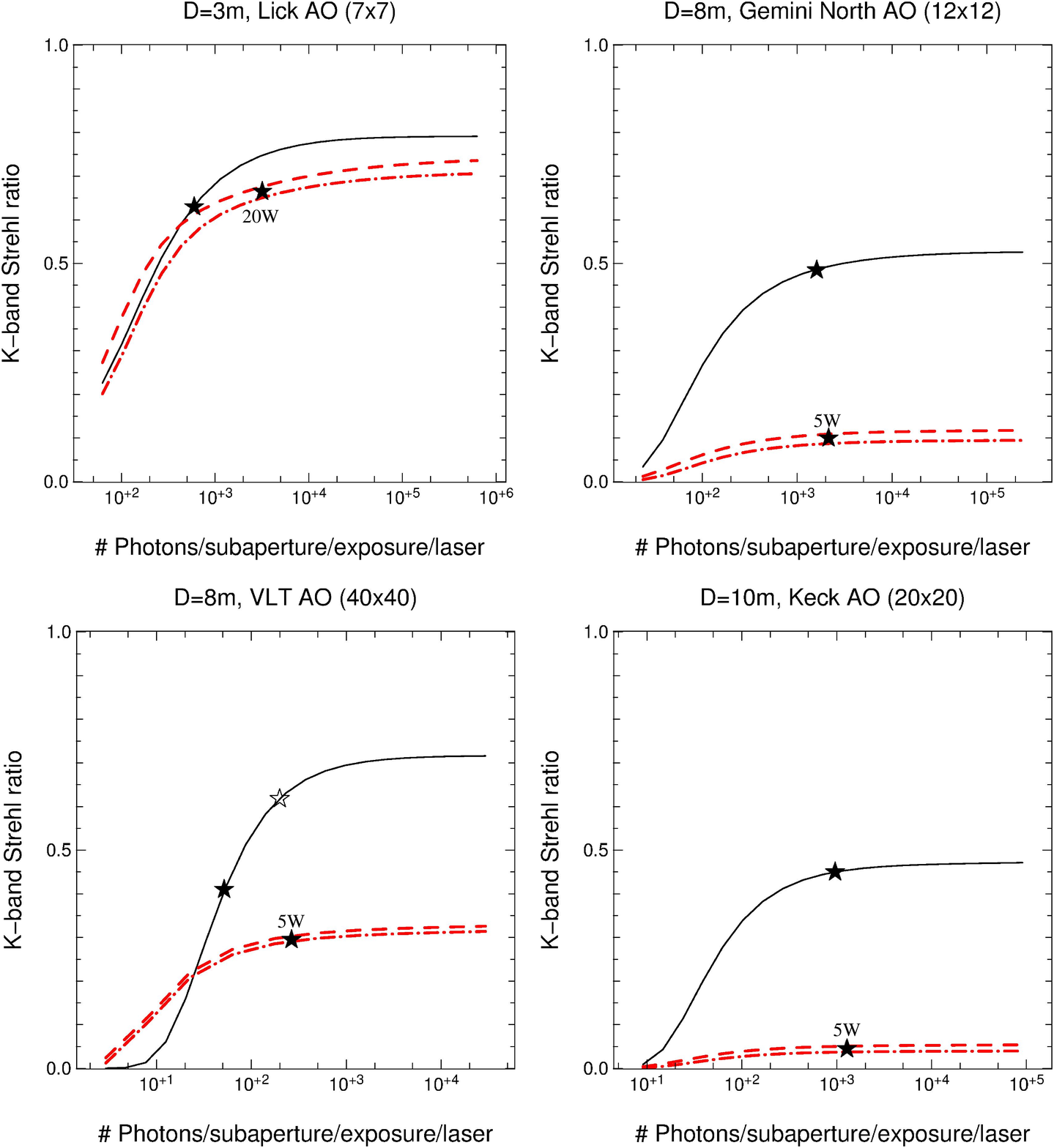}
\end{tabular}
\caption{\label{fig_perfsAO} K-band performance of one Sodium (solid lines) and 3/6 Rayleigh (dashdotted/dashed lines) laser stars as a function of the photon flux. In the Sodium case the star indicates the expected return flux for a 15W laser. The filled star in the case of the VLT shows the current operating point according to  Wizinovitch \cite{wizinowich_2}. For the Rayleigh case, the stars shows the minimum power required for reaching the saturation regime, varying from 5W to 20W.}  
\end{center}
\end{figure*}
Sodium LGS have been proven to provide a better correction than compared to Rayleigh LGS because of a much less severe cone effect. However, as Sodium lasers are 
substentially more expensive by a factor of $\gsim 10$ and may even dominate the cost of the full AO system, it is interesting to compare the quality of AO correction between \textit{a single} Sodium LGS and \textit{several}  Rayleigh LGSs, as long as the overall cost of the AO remains smaller in the latter case. We investigate this trade-off for different classes of telescope diameter in the context of existing observatories where Sodium laser devices have been installed and can be used for their AO system, namely: Lick ($d=3$m, $7\times7$ subapertures), Gemini North ($d=8$m, $12\times12$), VLT ($d=8$m, $40\times40$) and Keck ($d=8$m, $20\times20$) observatories. \\
We have used experimental data obtained with combined MASS-DIMM site testing instruments \cite{kornilov_1} to estimate atmospheric conditions above Mauna Kea Observatory (Gemini, Keck). It consists in 6 layers located at $[0,1,2,4,8,16]$km, with relative $C_n^2$ contributions of $[53,11,4,12,9,11]\%$. For Mount Hamilton (Lick) and Paranal (VLT) observatories we have used the theoretical Hufnagel night profile of Sect. (\ref{sec_el}). Both profiles have been generated for the same average seeing conditions (i.e. $r_0=12$cm in R band) although the Hufnagel $C_n^2$ is probably leading to more optimistic results as the contribution of the upper layers is lower in this case than that of the measurements at Mauna Kea. The numbers of SH subapertures of the wavefront sensors correspond to the actual AO instruments in operation, as summarized by Wizinovitch \cite{wizinowich_2}. We have chosen an average AO bandwidth of $250$Hz and an associated loop closing time of $\tau=4$ms. Finally, we have set the detector noise to $\sigma_{d}=1e^{-}$. Figure (\ref{fig_perfsAO}) displays the K-band Strehl ratio as a function of the incoming number of photons per subaperture, covering the different noise regimes as the photon flux increases: detector noise, photon noise and fitting error/cone effect plateau, respectively. For the Sodium LGS case, we have indicated the expected return photon flux of a $15$W laser with a star symbol. From the lidar equation  \cite{hardy_1}, it corresponds roughly to a spot brightness of $N \sim 1.5\times 10^6\mathrm{ph}/\mathrm{m}^2/\mathrm{s}$ or equivalently to a star magnitude of $V\simeq 9.5$. These numbers are consistent with the properties of the laser effectively used for Lick, Gemini, and Keck \cite{wizinowich_2}. For the VLT, while the specification requires a return flux of $N \ge 1 \times 10^6\mathrm{ph}/\mathrm{m}^2/\mathrm{s}$ \cite{bonaccini_2}, it seems that the actual laser rather provides a $V \simeq 11$ artificial spot \cite{wizinowich_2}. Both options ($V \simeq 11$ and $V \simeq 9.5$) are reported in this case. Alternatively, Rayleigh star return flux is indicated considering the minimum power required to reach optimal performance (saturation regime). We find that a minimum power of $\sim 5$W$-20$W is needed, values that are main stream numbers for that class of lasers \cite{hardy_1}. We note that the contribution of LGS AO (i.e. without tip/tilt error) to the error budget obtained from Keck science images \cite{van_dam_1} (see Table 1. in their paper) gives a K-band Strehl ratio of $\sim 0.5$, which is in good agreement with our theoretical estimations ($\sim 0.45$). Similarly, the estimated (LGS/AO) K-band Strehl ratio of $\sim 0.7$ computed by Max \textit{et al.} \cite{max_1} for the Lick Sodium LGS (see Table 1. in their paper) is consistent with our predictions ($\sim 0.6$).\\
We however emphasize that the present Sodium laser operating points are at the very edge or even below the plateau region that represents the maximum achievable performance. Although the estimation of the return flux is of debate since it will strongly depend on various factors like the Sodium abundance in the mesospheric layer, we assert that more powerful lasers are quite likely to improve the performance of LGS AO correction of these observatories. The improvment would be significant especially for Lick and VLT telescopes with a potential K-band Strehl ratio increase of $\sim 15\%$. It would however require lasers with power $2$ to $5$ times stronger than those in operation, hence driving to a substential growth of the cost of the instrument. \\
In the case of Lick telescope, we can see that Rayleigh stars can represent a very interesting alternative since only 3 such lasers will allow reaching performance equivalent to that of the present Sodium LGS.  We therefore stress that, for the $\lsim 5$m class telescopes, this approach may offer an excellent potential in terms of benefits/cost. On the contrary, the situation severly shifts in favour of Sodium LGS when the telescope size increases, as the cone effect becomes too strong to be compensated by a network of several Rayleigh stars, as infered by Le Louarn \textit{et al.} \cite{lelouarn_1}. Even doubling the number of Rayleigh stars from $3$ to $6$ is far from reaching the performance of Sodium LGS. As a consequence, the Rayleigh LGS solution will not give the same performance as that of a Sodium LGS for the $\gsim 8$m class telescopes. However, if cost is a prority driver, for a loss in K-band Strehl ratio of $\sim 50\%$, one could get a mulit-Rayleigh LGS at a fraction of the cost of a Sodium LGS. In an even more drastic way, Extremely Large Telescopes (ELT) of diameters $\gsim 30$m, that are contemplated to be be built in the next decade, will absolutely be unable to work with Rayleigh laser guide systems.  

\section{Conclusion}
We have provided in this paper an analytical derivation of the performance of LTAO technique, demonstrating that the phase residual error can be formally described by a combination of integrals of product of three Bessel functions. Thanks to this formalism, we have quantified the limitations of AO performance arising from the combined effect of partial wavefront sensing, time delay and cone effect when using one LGS. The latter effect can be fought by using several guide stars and performing a tomographic reconstruction of the turbulent volume. In the case of Sodium lasers, the compensation of focal anisoplanatism can be total with a moderate number ($\ge 3$) of artificial spots evenly distributed in the sky on a circle of angular radius $0.5D/L$. With Rayleigh stars, for which cone effect is much stronger, focal anisoplanatism can be only partially corrected, even when using a great number of beacons, because the upper turbulent layers cannot be fully mapped by the laser beams. This fundamental limitation has often led to consider Rayleigh stars unsuitable for astronomical purposes.  However, when dealing with small diameter class telescopes ($\lsim 5$m), using a few ($\sim 3$) of such lasers instead of a single Sodium one should be considered as a conceivable alternative for it can provide equivalent AO correction with a lower overall cost of the instrument.

\section*{Acknowledgements}
We would like to thank Dr. Warren Skidmore for providing the experimental turbulent profiles of Mauna Kea observatory and for his useful comments about the methods used to obtain these measurements.

\newpage
\appendix
\renewcommand{\theequation}{A-\arabic{equation}}
% redefine the command that creates the equation no.
\setcounter{equation}{0}  % reset counter \appendix
\renewcommand{\thetable}{A-\arabic{table}}
\setcounter{table}{0}  % reset counter \appendix

\section*{Appendix A: Bessel and Zernike functions properties}
\subsection*{A.1 Integral forms of Bessel functions}\label{app_bess}
We recall the properties of the $J_m$ Bessel functions in their integral forms. They will be used to derive the equations of following appendices:
\begin{eqnarray}
&&\int_0^{2\pi} \cos(m\gamma)\exp(iy\cos(\gamma-\theta_k)) \rm{d}\gamma = \left\{\begin{array}{l}2\pi(-1)^\frac{|m|}{2}\cos(m\theta_k)J_{|m|}(y)~~ \mathrm{if~}m~\mathrm{even} \\ 2i\pi(-1)^\frac{|m|-1}{2}\cos(m\theta_k)J_{|m|}(y)~~ \mathrm{if~}m~\mathrm{odd}\end{array}\right. \label{eq_jm_cos}\\
&&\int_0^{2\pi} \sin(m\gamma)\exp(iy\cos(\gamma-\theta_k)) \rm{d}\gamma = \left\{\begin{array}{l}2\pi(-1)^\frac{|m|}{2}\sin(m\theta_k)J_{|m|}(y)~~ \mathrm{if~}m~\mathrm{even} \\ 2i\pi(-1)^\frac{|m|-1}{2}\sin(m\theta_k)J_{|m|}(y)~~ \mathrm{if~}m~\mathrm{odd}\end{array}\right. \label{eq_jm_sin}
\end{eqnarray}
\subsection*{A.2. Zernike polynomials characteristics}\label{app_zern}
In polar coordinates. the Zernike modes are defined for a circular aperture without obstruction as:
\begin{equation}
Z^m_n(\rho, \theta) = Z_j(\rho, \theta) = \sqrt{n+1}R^m_n(\rho) \left\{\begin{array}{l}\sqrt{2}\cos(|m|\theta)~~\mathrm{if}~m>0\\ \sqrt{2}\sin(|m|\theta)~~\mathrm{if}~m<0\\1~~\mathrm{if}~m=0\end{array}\right.
\end{equation}
where $n$ and $m$ are respectively the radial degree and the azimuthal frequency of the $j^{th}$ polynomial, $j$ being defined as $j=\frac{n(n+2)+m}{2}$, and:
 \begin{equation}
R^m_n(\rho)  = \sum_{s=0}^{(n-|m|)/2} \frac{(-1)^s(n-s)!}{s![(n+|m|)/2-s]![(n-|m|)/2-s]!}\rho^{n-2s}
\end{equation}
The Zernike modes are orthonormal over a circle of unit radius, that is: 
\begin{equation}
\int \Pi_p(\ro)Z_j(\ro)Z_k(\ro) \mathrm{d}^2\ro =   \left\{\begin{array}{l} 1~~\mathrm{if}~j = k\\ 0~~\mathrm{if}~j \neq k\end{array}\right.
\end{equation}
with $\Pi_p(\ro)$ being the unitary pupil function.\\
For a given phase $\Phi(R\ro)$ defined over a pupil of radius $R$, its Zernike decomposition is expressed as $\Phi(R\ro) = \sum_{j=0}^{\infty} \phi_j Z_j(\ro)$, where the Zernike coefficients are calculated by projecting the phase on the polynomial basis:
\begin{equation}
\phi_j = \int \Pi_p(\ro)Z_j(\ro)\Phi(R\ro)\mathrm{d}^2\ro \label{eq_ai}
\end{equation} 
$Q_j(\kap)$, the Fourier Transform of $\Pi_p(\ro)Z_j(\ro)$, can be written as:
\begin{equation}
Q_j(\kappa, \gamma) = (-1)^n\sqrt{n+1}\frac{J_{n+1}(2\pi\kappa)}{\pi\kappa}\left\{\begin{array}{l}(-1)^{(n-|m|)/2}i^{|m|}\sqrt{2}\cos(|m|\gamma)~~\mathrm{if}~m>0\\ (-1)^{(n-|m|)/2}i^{|m|}\sqrt{2}\sin(|m|\gamma)~~\mathrm{if}~m<0\\(-1)^{n/2}~~\mathrm{if}~m=0\end{array}\right. \label{eq_tfzern}
\end{equation}
\subsection*{A.3. The elements of the interaction matrix}\label{app_interactionmat}
The elements of the interaction matrix $D_{\infty}$ are defined in Eq. (\ref{eq_system}) and can be rewritten as following:
\begin{equation}
D^{x,y}_{kj} = \frac{\lambda{R}}{2\pi A_s} \int \pi \frac{\partial [\Pi_p(\ro)Z_j(\ro)]}{\partial{x,y}} \Pi_s^k\left(\frac{R}{R_s}\ro\right) \rm{d}^2\ro
\end{equation}
By making use of Fourier Transform, the previous equation becomes:
\begin{equation}
D^{x,y}_{kj} = \frac{\lambda{R}}{2\pi A_s} \int \pi.2i\pi\kappa_{x,y} Q_j(\kap)\widehat{\Pi_s^k}(-\kap) \rm{d}^2\kap 
\end{equation}
Assuming circular subapertures, that is $\widehat{\Pi_s^k}(\kap)$ as in Eq. (\ref{eq_ftpis}), we obtain:
\begin{eqnarray}
D^{x,y}_{kj}& =&  \frac{i\lambda{R}\pi 2\pi}{2\pi A_s}\left[\frac{Rs}{R}\right]\int \kappa_{x,y} Q_j(\kap)\frac{J_1\left(2\pi\frac{R_s|\kap|}{R}\right)}{|\kap|}\exp^{-2i\pi\ro_k.\kap}\rm{d}^2\kap \nonumber \\
&=& \frac{i\lambda}{R_s}\int \kappa_{x,y} Q_j(\kap)\frac{J_1\left(2\pi\frac{R_s|\kap|}{R}\right)}{|\kap|}\exp^{2i\pi\ro_k.\kap}\rm{d}^2\kap
\end{eqnarray}
Switching to polar coordinates with $\ro_k=[\rho_k, \theta_k]$ and $\kap=[\kappa, \gamma]$, we have:
\begin{eqnarray}
&&\left[\begin{array}{cc} D_{kj}^{x} \\ D_{kj}^{y}\end{array}\right] = (-1)^n\frac{i\lambda}{\pi R_s}  \sqrt{n+1} \int_0^{\infty} \rm{d}\kappa~ J_1\left(2\pi\frac{R_s}{R}\kappa\right)J_{n+1}(2\pi\kappa)  \\
&&\times\int_0^{2\pi} \mathrm{d}\gamma~\left[\begin{array}{cc} \cos(\gamma) \\ \sin(\gamma)\end{array}\right] \exp^{2i\pi\rho_k\kappa\cos(\gamma-\theta_k)}\left\{\begin{array}{l}(-1)^{(n-|m|)/2}i^{|m|}\sqrt{2}\cos(|m|\gamma)~~\mathrm{if}~m>0\\ (-1)^{(n-|m|)/2}i^{|m|}\sqrt{2}\sin(|m|\gamma)~~\mathrm{if}~m<0\\(-1)^{n/2}~~\mathrm{if}~m=0\end{array}\right. \nonumber
\end{eqnarray}
In the integral over $\gamma$, we recognize the Bessel functions of App. (A.1) that we explicitly define in Table (\ref{table_bess}).
\begin{table*}
\caption{\label{table_bess} Evaluation of integrals in terms of Bessel functions.}
\begin{tabular}{|c|}
\hline
\begin{minipage}{\textwidth}
\begin{eqnarray}
&&\int_0^{2\pi} \mathrm{d}\gamma~\cos(\gamma)\cos(|m|\gamma)\exp^{iy\cos(\gamma-\theta_k)} =  \nonumber\\
&& \pi\left[\cos([|m|-1]\theta_k)J_{|m|-1}\left(y\right)- \cos([|m|+1]\theta_k)J_{|m|+1}\left(y\right)\right]  \times \left\{\begin{array}{r} i (-1)^\frac{|m|-2}{2}~~\mathrm{if}~|m|~\mathrm{even} \\ (-1)^\frac{|m|-1}{2}~~\mathrm{if}~|m|~\mathrm{odd}~\end{array}\right.\nonumber  \\
&&\int_0^{2\pi} \mathrm{d}\gamma~\cos(\gamma)\sin(|m|\gamma)\exp^{iy\cos(\gamma-\theta_k)} = \nonumber\\
&& \pi\left[\sin([|m|-1]\theta_k)J_{|m|-1}\left(y\right)- \sin([|m|+1]\theta_k)J_{|m|+1}\left(y\right)\right] \times \left\{\begin{array}{r} i (-1)^\frac{|m|-2}{2}~~\mathrm{if}~|m|~\mathrm{even} \\ (-1)^\frac{|m|-1}{2}~~\mathrm{if}~|m|~\mathrm{odd}~\end{array}\right.  \nonumber \\
&&\int_0^{2\pi} \mathrm{d}\gamma~\cos(\gamma)\exp^{iy\cos(\gamma-\theta_k)} = 2i\pi\cos(\theta_k)J_1(y) \nonumber
\end{eqnarray}
\end{minipage}\\
\hline
\begin{minipage}{\textwidth}
\begin{eqnarray}
&&\int_0^{2\pi} \mathrm{d}\gamma~\sin(\gamma)\cos(|m|\gamma)\exp^{iy\cos(\gamma-\theta_k)} = \nonumber\\
&&\pi\left[\sin([|m|-1]\theta_k)J_{|m|-1}\left(y\right)+ \sin([|m|+1]\theta_k)J_{|m|+1}\left(y\right)\right] \times \left\{\begin{array}{r} i (-1)^\frac{|m|}{2}~~\mathrm{if}~|m|~\mathrm{even} \\ (-1)^\frac{|m|+1}{2}~~\mathrm{if}~|m|~\mathrm{odd}~\end{array}\right. \nonumber  \\
&&\int_0^{2\pi} \mathrm{d}\gamma~\sin(\gamma)\sin(|m|\gamma)\exp^{iy\cos(\gamma-\theta_k)} = \nonumber\\
&&-\pi \left[\cos([|m|-1]\theta_k)J_{|m|-1}\left(y\right)+ \cos([|m|+1]\theta_k)J_{|m|+1}\left(y\right)\right] \times \left\{\begin{array}{r} i (-1)^\frac{|m|}{2}~~\mathrm{if}~|m|~\mathrm{even} \\ (-1)^\frac{|m|+1}{2}~~\mathrm{if}~|m|~\mathrm{odd}~\end{array}\right.\nonumber  \\
&&\int_0^{2\pi} \mathrm{d}\gamma~\sin(\gamma)\exp^{iy\cos(\gamma-\theta_k)} = 2i\pi\sin(\theta_k)J_1(y) \nonumber
\end{eqnarray}
\end{minipage}\\
\hline
\end{tabular}
\end{table*}
This leads to the below expressions of the interaction matrix coefficients:
\begin{eqnarray}
&&D^{x}_{kj} =  \frac{\lambda}{R_s} s_{n,m}  \\
&&\times \int_0^{\infty} \rm{d}\kappa~ J_1\left(2\pi\frac{R_s}{R}\kappa\right)J_{n+1}(2\pi\kappa)\left[\beta^x_{|m|-1,k}J_{|m|-1}\left(2\pi\rho_k\kappa\right)- \beta^x_{|m|+1,k}J_{|m|+1}\left(2\pi\rho_k\kappa\right)\right]\nonumber\\
&&D^{y}_{kj} =  \frac{\lambda}{R_s} s_{n,m} \\
&&\times \int_0^{\infty} \rm{d}\kappa~ J_1\left(2\pi\frac{R_s}{R}\kappa\right)J_{n+1}(2\pi\kappa)\left[\beta^y_{|m|-1,k}J_{|m|-1}\left(2\pi\rho_k\kappa\right)+\beta^y_{|m|+1,k}J_{|m|+1}\left(2\pi\rho_k\kappa\right)\right] \nonumber\\
&& \mathrm{with} \nonumber \\
&&s_{n,m} =  i^{|m|}(-1)^\frac{3n}{2}\sqrt{n+1}\left\{\begin{array}{ll} \sqrt{2}&\mathrm{if}~m \ne 0 \\ 1&\mathrm{if}~m=0\end{array}\right.;\label{eq_snm_app} \\
&&\beta^x_{|m| \pm 1,k} = \left\{\begin{array}{ll} \cos([|m| \pm 1]\theta_k)&\mathrm{if}~m \ge 0\\ \sin([|m| \pm 1]\theta_k)&\mathrm{if}~m \le 0 \end{array}\right. \nonumber
\end{eqnarray}

\section*{Appendix B: Formal derivation of $\mathrm{Cov}(\vs)$}\label{app_covs}
\subsection*{B.1. Computation of $<s_k^{x,y}(\valpha_p)s_l^{x,y}(\valpha_q)>$}
Using Eqs. (\ref{eq_alpha}) we have:
\begin{equation}
<s_k^x(\valpha_p)s_l^x(\valpha_q)>=\left(\frac{\lambda{R}}{2\pi A_s}\right)^2
\iint_{subap_{(k,l)}}<\frac{\partial}{\partial{x_1}}[\Phi^{lgs}(R\ro_1,\valpha_p)]\frac{\partial}{\partial{x_2}}[\Phi^{lgs}(R\ro_2,\valpha_q)]> \mathrm{d}^2\ro_1\mathrm{d}^2\ro_2
\end{equation}
Introducing the subaperture function $\Pi_s$ and the LGS covariance matrix of Eq. (\ref{eq_covlgs}), we get:
\begin{eqnarray}
<s_k^x(\valpha_p)s_l^x(\valpha_q)> &=& \left(\frac{\lambda{R}}{2\pi A_s}\right)^2  \\
&\times& \iint \Pi_s^k\left(\frac{R}{R_s}\ro_1\right) \Pi_s^l\left(\frac{R}{R_s}\ro_2\right) \frac{\partial^2}{\partial{x_1}\partial{x_2}}B_{\Phi}^{lgs}(R[\ro_1-\ro_2], \Delta\valpha_{qp})\mathrm{d}^2\ro_1\mathrm{d}^2\ro_2 \nonumber \\
&=& -\left(\frac{\lambda{R^2}}{2\pi A_s}\right)^2 \iint \mathrm{d}^2\ro_1\mathrm{d}^2\ro_2 \Pi_s^k\left(\frac{R}{R_s}\ro_1\right) \Pi_s^l\left(\frac{R}{R_s}\ro_2\right) \nonumber \\
&&\times \int^{L}_0[\zeta(h)]^2\frac{\partial^2 B_{\Delta{n}}^{h}}{\partial{x_1}\partial{x_2}}(\zeta(h)R[\ro_1-\ro_2]+h\Delta\valpha_{qp})\mathrm{d}{h}
\end{eqnarray}
where $ \Delta\valpha_{qp} = \valpha_p - \valpha_q$. Further, we follow the analytical development of Molodij \cite{molodij_1}, with the intermediate change of variable $\veta=\ro_1-\ro_2$. In our case however, we also take into account the derivative properties of Fourier Transform. Thus we have:
\begin{eqnarray}
<s_k^x(\valpha_p)s_l^x(\valpha_q)>  &=& \left(\frac{\lambda}{A_s}\right)^2 \int \mathrm{d}^2\kap~\widehat{\Pi_s^k}(\kap)\widehat{\Pi_s^l}^{\ast}(\kap)  \kappa_x^2  \nonumber \\
&&  \times \int_0^L \mathrm{d}{h}~\frac{1}{[\zeta(h)]^{2}}  W^{h}_{\Delta{n}}\left(-\frac{\kap}{R\zeta(h)}\right) \exp^{-2i\pi\frac{h\Delta\valpha_{qp}}{R\zeta(h)}.\kap} 
\end{eqnarray}
Using the definition of $ W^{h}_{\Delta{n}}(\kap)$ in Eq. (\ref{eq_wn}), $<s_k^x(\valpha_p)s_l^x(\valpha_q)>$ takes the generic form:
\begin{eqnarray}
<s_k^x(\valpha_p)s_l^x(\valpha_q)>  &=& \frac{0.023}{2^{\frac{5}{3}}\int_0^{\infty}C^2_n(h) \mathrm{d}{h}}\left(\frac{\lambda{R}}{A_s}\right)^2\left(\frac{D}{r_0}\right)^{\frac{5}{3}}   \nonumber \\
&&  \times  \int \mathrm{d}^2\kap~\widehat{\Pi_s^k}(\kap)\widehat{\Pi_s^l}^{\ast}(\kap)  \kappa_x^2|\kap|^{-\frac{11}{3}}  \nonumber \\
&&  \times \int_0^L \mathrm{d}{h}~[\zeta(h)]^\frac{5}{3} C^2_n(h) \exp^{-2i\pi\frac{h\Delta\valpha_{qp}}{R\zeta(h)}.\kap}   
\end{eqnarray}
Switching to polar coordinates and assuming circular subapertures, we get:
\begin{eqnarray}
<s_k^x(\valpha_p)s_l^x(\valpha_q)>  &=& \frac{0.023}{\pi^22^{\frac{5}{3}}\int_0^{\infty}C^2_n(h) \mathrm{d}{h}} \left(\frac{\lambda}{R_s}\right)^2 \left(\frac{D}{r_0}\right)^{\frac{5}{3}} \int_0^L \mathrm{d}{h} [\zeta(h)]^{\frac{5}{3}}C^2_n(h) \nonumber \\
&&  \times \int_0^{\infty} \rm{d}\kappa   \left[J_1\left(2\pi\frac{R_s}{R}\kappa\right)\right]^2  \kappa^{-\frac{8}{3}} \nonumber \\
&&  \times \int_0^{2\pi} \rm{d}\gamma \cos^2(\gamma)\exp^{2i\pi\rho^{pq}_{kl}(h)\kappa\cos(\gamma-\theta^{pq}_{kl}(h))}
\end{eqnarray}
where $\rho^{pq}_{kl}(h)$ and $\theta_{kl}^{pq}(h)$ are the modulus and the argument of the vector $\ro_l-\ro_k + \frac{h}{R\zeta(h)}\Delta\valpha_{pq}$, respectively.
For the integral over $\gamma$, once we rewrite $\cos^2(\gamma)$ as $\frac{1+\cos(2\gamma)}{2}$ we recognize the integral forms of Bessel functions of Table (\ref{table_bess}):
\begin{equation}
 \int_0^{2\pi} \rm{d}\gamma \cos^2(\gamma)\exp^{2i\pi\rho^{pq}_{kl}(h)\kappa\cos(\gamma-\theta^{pq}_{kl}(h))} = \pi[J_0\left(2\pi\rho^{pq}_{kl}(h)\kappa\right)-\cos(2\theta^{pq}_{kl}(h))J_2\left(2\pi\rho^{pq}_{kl}(h)\kappa\right)]
\end{equation}
We thus obtain the final expression for the moment $<s_k^x(\valpha_p)s_l^x(\valpha_q)>$ as an integral of product of Bessel functions:
\begin{eqnarray}
&&<s_k^x(\valpha_p)s_l^x(\valpha_q)>  = \frac{0.023}{\pi 2^{\frac{5}{3}}\int_0^{\infty}C^2_n(h) \mathrm{d}{h}} \left(\frac{\lambda}{R_s}\right)^2\left(\frac{D}{r_0}\right)^{\frac{5}{3}}\int_0^L  \mathrm{d}{h}~\left[\zeta(h)\right]^{\frac{5}{3}} C^2_n(h)\\
&& \hskip25pt \times   \int_0^{\infty}\mathrm{d}\kappa~\left[J_1\left(2\pi\frac{R_s}{R}\kappa\right)\right]^2 \kappa^{-\frac{8}{3}} [J_0\left(2\pi\rho^{pq}_{kl}(h)\kappa\right) - \cos(2\theta^{pq}_{kl}(h))\ J_2\left(2\pi\rho^{pq}_{kl}(h)\kappa\right)] \nonumber
\end{eqnarray}
The expression of associated moments $<s_k^y(\valpha_p)s_l^y(\valpha_q)>$ and $<s_k^x(\valpha_p)s_l^y(\valpha_q)>$ can be derived in a straighforward way by simple analogy: 
\begin{eqnarray}
<s_k^y(\valpha_p)s_l^y(\valpha_q)>  &=& \frac{0.023}{2^{\frac{5}{3}}\int_0^{\infty}C^2_n(h) \mathrm{d}{h}}\left(\frac{\lambda{R}}{A_s}\right)^2\left(\frac{D}{r_0}\right)^{\frac{5}{3}}   \nonumber \\
&&  \times  \int \mathrm{d}^2\kap~\widehat{\Pi_s^k}(\kap)\widehat{\Pi_s^l}^{\ast}(\kap)  \kappa_y^2|\kap|^{-\frac{11}{3}}  \nonumber \\
&&  \times \int_0^L \mathrm{d}{h}~\zeta(h)^\frac{5}{3} C^2_n(h) \exp^{-2i\pi\frac{h\Delta\valpha_{qp}}{R\zeta(h)}.\kap} 
\end{eqnarray}
\begin{eqnarray}
<s_k^x(\valpha_p)s_l^y(\valpha_q)>  &=& \frac{0.023}{2^{\frac{5}{3}}\int_0^{\infty}C^2_n(h) \mathrm{d}{h}}\left(\frac{\lambda{R}}{A_s}\right)^2\left(\frac{D}{r_0}\right)^{\frac{5}{3}}   \nonumber \\
&&  \times  \int \mathrm{d}^2\kap~\widehat{\Pi_s^k}(\kap)\widehat{\Pi_s^l}^{\ast}(\kap)  \kappa_x\kappa_y|\kap|^{-\frac{11}{3}}  \nonumber \\
&&  \times \int_0^L \mathrm{d}{h}~\zeta(h)^\frac{5}{3} C^2_n(h) \exp^{-2i\pi\frac{h\Delta\valpha_{qp}}{R\zeta(h)}.\kap}   
\end{eqnarray}
which, for circular subapertures become:
\begin{eqnarray}
&&<s_k^y(\valpha_p)s_l^y(\valpha_q)>  = \frac{0.023}{\pi 2^{\frac{5}{3}}\int_0^{\infty}C^2_n(h) \mathrm{d}{h}} \left(\frac{\lambda}{R_s}\right)^2\left(\frac{D}{r_0}\right)^{\frac{5}{3}}\int_0^L \mathrm{d}{h}~\left[\zeta(h)\right]^{\frac{5}{3}} C^2_n(h)\\
&& \hskip25pt \times  \int_0^{\infty} \mathrm{d}\kappa~\left[J_1\left(2\pi\frac{R_s}{R}\kappa\right)\right]^2 \kappa^{-\frac{8}{3}} [J_0\left(2\pi\rho^{pq}_{kl}(h)\kappa\right) + \cos(2\theta^{pq}_{kl}(h))\ J_2\left(2\pi\rho^{pq}_{kl}(h)\kappa\right)]   \nonumber\\
&&<s_k^x(\valpha_p)s_l^y(\valpha_q)>  = \frac{0.023}{\pi 2^{\frac{5}{3}}\int_0^{\infty}C^2_n(h) \mathrm{d}{h}} \left(\frac{\lambda}{R_s}\right)^2\left(\frac{D}{r_0}\right)^{\frac{5}{3}}\int_0^L \mathrm{d}{h}~\left[\zeta(h)\right]^{\frac{5}{3}} C^2_n(h)\\
&& \hskip25pt \times   \int_0^{\infty} \mathrm{d}\kappa~\left[J_1\left(2\pi\frac{R_s}{R}\kappa\right)\right]^2 \kappa^{-\frac{8}{3}} [-\sin(2\theta^{pq}_{kl}(h))\ J_2\left(2\pi\rho^{pq}_{kl}(h)\kappa\right)]   \nonumber
\end{eqnarray}

\subsection*{B.2. Computation of $<\phi^{lgs}_{1,2}(\valpha_p)\phi^{lgs}_{1,2}(\valpha_q)>$}
From the definition of Zernike tip/tilt coefficients of Eq. (\ref{eq_ai}) we have:
\begin{equation}
<\phi^{lgs}_1(\valpha_p)\phi^{lgs}_1(\valpha_q)> = \iint \pi_p(\ro_1)Z_1(\ro_1)\pi_p(\ro_2)Z_1(\ro_2)B_{\Phi}^{lgs}(R[\ro_1-\ro_2], \Delta\valpha_{qp}) \mathrm{d}^2\ro_1\mathrm{d}^2\ro_2
\end{equation}
Again, by my means of Fourier Transform properties and variable changes of Molodij \cite{molodij_1}, the previous equation changes to:
\begin{eqnarray}
<\phi^{lgs}_1(\valpha_p)\phi^{lgs}_1(\valpha_q)> &=& \frac{1}{R^2}\int \mathrm{d}^2\kap~Q_1(\kap)Q_1^{\ast}(\kap)\nonumber \\
&&  \times \int_0^L \mathrm{d}{h}~[\zeta(h)]^{-2}  W^{h}_{\Delta{n}}\left(-\frac{\kap}{R\zeta(h)}\right) \exp^{-2i\pi\frac{h\Delta\valpha_{qp}}{R\zeta(h)}.\kap} \\
&=& \frac{0.023}{2^{\frac{5}{3}}\int_0^{\infty}C^2_n(h) \mathrm{d}{h}} \left(\frac{D}{r_0}\right)^{\frac{5}{3}} \int \mathrm{d}^2\kap~Q_1(\kap)Q_1^{\ast}(\kap) |\kap|^{-\frac{11}{3}}\nonumber \\
&&  \times \int_0^L \mathrm{d}{h} \left[\zeta(h)\right]^{\frac{5}{3}} C^2_n(h)\exp^{-2i\pi\frac{h\Delta\valpha_{qp}}{R\zeta(h)}.\kap}   
\end{eqnarray}
From the expression of $Q_1$ using Eq. (\ref{eq_tfzern}), we develop the equation in polar coordinates:
\begin{eqnarray}
<\phi^{lgs}_1(\valpha_p)\phi^{lgs}_1(\valpha_q)> &=& \frac{4 \times 0.023}{\pi^22^{\frac{5}{3}}\int_0^{\infty}C^2_n(h) \mathrm{d}{h}} \left(\frac{D}{r_0}\right)^{\frac{5}{3}} \int_0^L \mathrm{d}{h} \left[\zeta(h)\right]^{\frac{5}{3}} C^2_n(h) \nonumber \\
&& \times \int_0^{\infty} \mathrm{d}\kappa \left[J_2(2\pi\kappa)\right]^2\kappa^{-\frac{14}{3}} \nonumber \\
&&  \times \int_0^{2\pi} \mathrm{d}\gamma \cos^2(\gamma)\exp^{2i\pi\rho^{pq}(h)\kappa\cos(\gamma-\theta^{pq})} 
\end{eqnarray}
where $\rho^{pq}(h)$ and $\theta^{pq}$ are the modulus and the argument of $\frac{h\Delta\valpha_{pq}}{R\zeta(h)}$. According to the integral definition of Bessel functions, we obtain:
\begin{eqnarray}
&&<\phi^{lgs}_1(\valpha_p)\phi^{lgs}_1(\valpha_q)> = \frac{4 \times 0.023}{\pi 2^{\frac{5}{3}}\int_0^{\infty}C^2_n(h) \mathrm{d}{h}} \left(\frac{D}{r_0}\right)^{\frac{5}{3}} \int_0^L \mathrm{d}{h} \left[\zeta(h)\right]^{\frac{5}{3}} C^2_n(h) \\
&& \hskip25pt \times \int_0^{\infty} \mathrm{d}\kappa \left[J_2(2\pi\kappa)\right]^2\kappa^{-\frac{14}{3}} [J_0\left(2\pi\rho^{pq}(h)\kappa\right)-\cos(2\theta^{pq}(h))J_2\left(2\pi\rho^{pq}(h)\kappa\right)]  \nonumber
\end{eqnarray}
Similarly, from the definition of $Q_2$ relative to the tilt Zernike coefficient $\phi^{lgs}_2$, we obtain:
\begin{eqnarray}
&&<\phi^{lgs}_2(\valpha_p)\phi^{lgs}_2(\valpha_q)> = \frac{4 \times 0.023}{\pi 2^{\frac{5}{3}}\int_0^{\infty}C^2_n(h) \mathrm{d}{h}} \left(\frac{D}{r_0}\right)^{\frac{5}{3}} \int_0^L \mathrm{d}{h} \left[\zeta(h)\right]^{\frac{5}{3}} C^2_n(h) \\
&& \hskip25pt \times \int_0^{\infty} \mathrm{d}\kappa \left[J_2(2\pi\kappa)\right]^2\kappa^{-\frac{14}{3}} [J_0\left(2\pi\rho^{pq}(h)\kappa\right)+\cos(2\theta^{pq}(h))J_2\left(2\pi\rho^{pq}(h)\kappa\right)]  \nonumber \\
&&<\phi^{lgs}_1(\valpha_p)\phi^{lgs}_2(\valpha_q)> = \frac{4 \times 0.023}{\pi 2^{\frac{5}{3}}\int_0^{\infty}C^2_n(h) \mathrm{d}{h}} \left(\frac{D}{r_0}\right)^{\frac{5}{3}} \int_0^L \mathrm{d}{h} \left[\zeta(h)\right]^{\frac{5}{3}} C^2_n(h) \\
&& \hskip25pt \times \int_0^{\infty} \mathrm{d}\kappa \left[J_2(2\pi\kappa)\right]^2\kappa^{-\frac{14}{3}} [-\sin(2\theta^{pq}(h))J_2\left(2\pi\rho^{pq}(h)\kappa\right)]  \nonumber
\end{eqnarray}

\subsection*{B.3. Computation of $<s_k^{x,y}(\valpha_p)\phi^{lgs}_{1,2}(\valpha_q)>$}
Combining Eq. (\ref{eq_alpha}) and Eq. (\ref{eq_ai}) we have:
\begin{eqnarray}
<s_k^x(\valpha_p)\phi_1^{lgs}(\valpha_q)>&=&\left(\frac{\lambda{R}}{2\pi A_s}\right) \\
&\times& \iint  \Pi_s^k\left(\frac{R}{R_s}\ro_1\right)\pi_p(\ro_2)Z_1(\ro_2)\frac{\partial}{\partial{x_1}}[B_{\Phi}^{lgs}(R[\ro_1-\ro_2], \Delta\valpha_{qp})]\mathrm{d}^2\ro_1\mathrm{d}^2\ro_2 \nonumber \\
&=& \left(\frac{\lambda{R^2}}{2\pi A_s}\right) \iint \mathrm{d}^2\ro_1\mathrm{d}^2\ro_2 \Pi_s^k\left(\frac{R}{R_s}\ro_1\right)\pi_p(\ro_2)Z_1(\ro_2) \nonumber \\
&&\times \int^{L}_0[\zeta(h)]\frac{\partial B_{\Delta{n}}^{h}}{\partial{x_1}}(\zeta(h)R[\ro_1-\ro_2]+h\Delta\valpha_{qp})\mathrm{d}{h}
\end{eqnarray}
which in the Fourier plane rewrites in the following generic form: 
\begin{eqnarray}
<s_k^x(\valpha_p)\phi_1^{lgs}(\valpha_q)>&=&-i\left(\frac{\lambda}{R A_s}\right) \int \mathrm{d}^2\kap~\widehat{\Pi_s^k}(\kap)Q^{\ast}_1(\kap)  \kappa_x  \nonumber \\
&&  \times \int_0^L \mathrm{d}{h}~\frac{1}{[\zeta(h)]^{2}}  W^{h}_{\Delta{n}}\left(-\frac{\kap}{R\zeta(h)}\right) \exp^{-2i\pi\frac{h\Delta\valpha_{qp}}{R\zeta(h)}.\kap} \\
 &=& -i\frac{0.023}{2^{\frac{5}{3}}\int_0^{\infty}C^2_n(h) \mathrm{d}{h}}\left(\frac{\lambda{R}}{A_s}\right)\left(\frac{D}{r_0}\right)^{\frac{5}{3}}   \nonumber \\
&&  \times  \int \mathrm{d}^2\kap~\widehat{\Pi_s^k}(\kap)Q_1^{\ast}(\kap)  \kappa_x|\kap|^{-\frac{11}{3}}  \nonumber \\
&&  \times \int_0^L \mathrm{d}{h}~[\zeta(h)]^\frac{5}{3} C^2_n(h) \exp^{-2i\pi\frac{h\Delta\valpha_{qp}}{R\zeta(h)}.\kap}   
\end{eqnarray}
Using Eq. (\ref{eq_tfzern}) and assuming circular subapertures, the previous equation becomes in polar coordinates: 
\begin{eqnarray}
<s_k^x(\valpha_p)\phi_1^{lgs}(\valpha_q)>&=&\frac{2\times0.023}{\pi^22^{\frac{5}{3}}\int_0^{\infty}C^2_n(h) \mathrm{d}{h}}\left(\frac{\lambda}{R_s}\right)\left(\frac{D}{r_0}\right)^{\frac{5}{3}}\int_0^L \mathrm{d}{h}~[\zeta(h)]^\frac{5}{3} C^2_n(h)\nonumber \\
&& \times \int_0^{\infty} \mathrm{d}\kappa~J_1\left(2\pi\frac{R_s}{R}\kappa\right)J_2(2\pi\kappa)\kappa^{-\frac{11}{3}} \nonumber \\
&&  \times \int_0^{2\pi} \mathrm{d}\gamma \cos^2(\gamma)\exp^{2i\pi\rho_k^{pq}(h)\kappa\cos(\gamma-\theta_k^{pq})} 
\end{eqnarray}
where $\rho_k^{pq}(h)$ and $\theta_k^{pq}$ are the modulus and the argument of $\frac{h\Delta\valpha_{pq}}{R\zeta(h)}-\rho_k$.
Again we introduce the integral definition of Bessel functions so that we finally obtain:
\begin{eqnarray}
&&<s_k^x(\valpha_p)\phi_1^{lgs}(\valpha_q)>=\frac{2\times0.023}{\pi 2^{\frac{5}{3}}\int_0^{\infty}C^2_n(h) \mathrm{d}{h}}\left(\frac{\lambda}{R_s}\right)\left(\frac{D}{r_0}\right)^{\frac{5}{3}}\int_0^L \mathrm{d}{h}~[\zeta(h)]^\frac{5}{3} C^2_n(h) \\
&& \hskip20pt \times \int_0^{\infty} \mathrm{d}\kappa~J_1\left(2\pi\frac{R_s}{R}\kappa\right)J_2(2\pi\kappa)\kappa^{-\frac{11}{3}}[J_0\left(2\pi\rho_k^{pq}(h)\kappa\right)-\cos(2\theta_k^{pq}(h))J_2\left(2\pi\rho^{pq}_{k}(h)\kappa\right)]\nonumber
\end{eqnarray}
By analogy, we compute the remaining moments:
\begin{eqnarray}
<s_k^y(\valpha_p)\phi_2^{lgs}(\valpha_q)>&=& -i\frac{0.023}{2^{\frac{5}{3}}\int_0^{\infty}C^2_n(h) \mathrm{d}{h}}\left(\frac{\lambda{R}}{A_s}\right)\left(\frac{D}{r_0}\right)^{\frac{5}{3}}  \nonumber \\
&&  \times  \int \mathrm{d}^2\kap~\widehat{\Pi_s^k}(\kap)Q_2^{\ast}(\kap)  \kappa_y|\kap|^{-\frac{11}{3}}  \nonumber \\
&&  \times \int_0^L \mathrm{d}{h}~[\zeta(h)]^\frac{5}{3} C^2_n(h) \exp^{-2i\pi\frac{h\Delta\valpha_{qp}}{R\zeta(h)}.\kap}   
\end{eqnarray}
\begin{eqnarray}
<s_k^x(\valpha_p)\phi_2^{lgs}(\valpha_q)>&=& -i\frac{0.023}{2^{\frac{5}{3}}\int_0^{\infty}C^2_n(h) \mathrm{d}{h}}\left(\frac{\lambda{R}}{A_s}\right)\left(\frac{D}{r_0}\right)^{\frac{5}{3}}  \nonumber \\
&&  \times  \int \mathrm{d}^2\kap~\widehat{\Pi_s^k}(\kap)Q_2^{\ast}(\kap)  \kappa_x|\kap|^{-\frac{11}{3}}  \nonumber \\
&&  \times \int_0^L \mathrm{d}{h}~[\zeta(h)]^\frac{5}{3} C^2_n(h) \exp^{-2i\pi\frac{h\Delta\valpha_{qp}}{R\zeta(h)}.\kap}   
\end{eqnarray}
\begin{eqnarray}
<s_k^y(\valpha_p)\phi_1^{lgs}(\valpha_q)>&=& -i\frac{0.023}{2^{\frac{5}{3}}\int_0^{\infty}C^2_n(h) \mathrm{d}{h}}\left(\frac{\lambda{R}}{A_s}\right)\left(\frac{D}{r_0}\right)^{\frac{5}{3}}  \nonumber \\
&&  \times  \int \mathrm{d}^2\kap~\widehat{\Pi_s^k}(\kap)Q_1^{\ast}(\kap)  \kappa_y|\kap|^{-\frac{11}{3}}  \nonumber \\
&&  \times \int_0^L \mathrm{d}{h}~[\zeta(h)]^\frac{5}{3} C^2_n(h) \exp^{-2i\pi\frac{h\Delta\valpha_{qp}}{R\zeta(h)}.\kap}   
\end{eqnarray}
which in polar coordinates and assuming circular subapertures gives:
\begin{eqnarray}
&&<s_k^y(\valpha_p)\phi_2^{lgs}(\valpha_q)>=\frac{2\times0.023}{\pi 2^{\frac{5}{3}}\int_0^{\infty}C^2_n(h) \mathrm{d}{h}}\left(\frac{\lambda}{R_s}\right)\left(\frac{D}{r_0}\right)^{\frac{5}{3}}\int_0^L \mathrm{d}{h}~[\zeta(h)]^\frac{5}{3} C^2_n(h) \\
&& \hskip20pt \times \int_0^{\infty} \mathrm{d}\kappa~J_1\left(2\pi\frac{R_s}{R}\kappa\right)J_2(2\pi\kappa)\kappa^{-\frac{11}{3}}[J_0\left(2\pi\rho_k^{pq}(h)\kappa\right)+\cos(2\theta_k^{pq}(h))J_2\left(2\pi\rho^{pq}_{k}(h)\kappa\right)]\nonumber
\end{eqnarray}
\begin{eqnarray}
&&<s_k^x(\valpha_p)\phi_2^{lgs}(\valpha_q)>=<s_k^y(\valpha_p)\phi_1^{lgs}(\valpha_q)>=
\frac{2\times0.023}{\pi 2^{\frac{5}{3}}\int_0^{\infty}C^2_n(h) \mathrm{d}{h}}\left(\frac{\lambda}{R_s}\right)\left(\frac{D}{r_0}\right)^{\frac{5}{3}} \\
&& \hskip20pt \times \int_0^L \mathrm{d}{h}~[\zeta(h)]^\frac{5}{3} C^2_n(h)\int_0^{\infty} \mathrm{d}\kappa~J_1\left(2\pi\frac{R_s}{R}\kappa\right)J_2(2\pi\kappa)\kappa^{-\frac{11}{3}}[-\sin(2\theta_k^{pq}(h))J_2\left(2\pi\rho^{pq}_{k}(h)\kappa\right)]\nonumber
\end{eqnarray}

\section*{Appendix C: Formal derivation of $\mathrm{Cov}(\vs,\vphi)$}\label{app_covphis}
\subsection*{C.1. Computation of $<s_k^{x,y}(\valpha_p)\phi_{j}>$}
Combining Eq. (\ref{eq_alpha}) and Eq. (\ref{eq_ai}) leads to:
\begin{eqnarray}
<s_k^x(\valpha_p)\phi_j>&=&\left(\frac{\lambda{R}}{2\pi A_s}\right)\iint \mathrm{d}^2\ro_1\mathrm{d}^2\ro_2~\Pi_s^k\left(\frac{R}{R_s}\ro_1\right)\pi_p(\ro_2)Z_j(\ro_2) \\
&&\times <\frac{\partial}{\partial{x_1}}[\phi^{lgs}(R\ro_1, \valpha_p)]\phi(R\ro_2)> \nonumber \\
&=& \left(\frac{\lambda{R^2}}{2\pi A_s}\right) \iint \mathrm{d}^2\ro_1\mathrm{d}^2\ro_2 \Pi_s^k\left(\frac{R}{R_s}\ro_1\right)\pi_p(\ro_2)Z_j(\ro_2) \nonumber \\
&&\times \int^{L}_0[\zeta(h)]\frac{\partial B_{\Delta{n}}^{h}}{\partial{x_1}}(R[\zeta(h)\ro_1-\ro_2]+h\valpha_{p})\mathrm{d}{h}
\end{eqnarray}
We perform the change of variable $\veta=\zeta(h)\ro_1-\ro_2$ \cite{molodij_1}, and we make use of the derivative properties of Fourier Transform to obtain the generic expression of the moment:
\begin{eqnarray}
<s_k^x(\valpha_p)\phi_j>&=&i\left(\frac{\lambda}{R A_s}\right) \int \mathrm{d}^2\kap~\int_0^L \mathrm{d}{h} \widehat{\Pi_s^k}^{\ast}(\zeta(h)\kap)Q_j(\kap)  \kappa_x [\zeta(h)] W^{h}_{\Delta{n}}\left(\frac{\kap}{R}\right) \exp^{2i\pi\frac{h\valpha_{p}}{R}.\kap}  \nonumber  \\
 &=& i\frac{0.023}{2^{\frac{5}{3}}\int_0^{\infty}C^2_n(h) \mathrm{d}{h}}\left(\frac{\lambda{R}}{A_s}\right)\left(\frac{D}{r_0}\right)^{\frac{5}{3}}   \nonumber \\
&&  \times  \int \mathrm{d}^2\kap \int_0^L \mathrm{d}{h}~\widehat{\Pi_s^k}^{\ast}(\zeta(h)\kap)Q_j(\kap)  \kappa_x|\kap|^{-\frac{11}{3}} [\zeta(h)] C^2_n(h) \exp^{2i\pi\frac{h\valpha_{p}}{R}.\kap}
\end{eqnarray}
Using the definitions of Zernike polynomials and circular subaperture Fourier Transform, the previous equation can be rewritten in polar coordinates as following:
\begin{eqnarray}
<s_k^x(\valpha_p)\phi_j>&=&  i(-1)^{-\frac{m}{2}}s_{n,m}\frac{0.023}{\pi^22^{\frac{5}{3}}\int_0^{\infty}C^2_n(h) \mathrm{d}{h}}\left(\frac{\lambda}{R_s}\right)\left(\frac{D}{r_0}\right)^{\frac{5}{3}}  \int_0^L \mathrm{d}{h}~C^2_n(h) \nonumber \\
&&\times \int_0^{\infty} \mathrm{d}\kappa~J_1\left(2\pi\zeta(h)\frac{R_s}{R}\kappa\right)J_2(2\pi\kappa)\kappa^{-\frac{11}{3}}  \nonumber\\
&&\times \int_0^{2\pi} \mathrm{d}\gamma \cos(\gamma)\left\{\begin{array}{c}\cos(|m|\gamma)\\\sin(|m|\gamma)\\1\end{array}\right\}\exp^{2i\pi\rho_k^{p}(h)\kappa\cos(\gamma-\theta_k^{p}(h))} 
\end{eqnarray}
where $\rho_k^{p}(h)$ and $\theta_k^{p}(h)$ are the modulus and the argument of $\frac{h\valpha_{p}}{R}+\zeta(h)\rho_k$ and $s_{n,m}$ is defined by Eq. (\ref{eq_snm}). The different cases of the integral over $\gamma$ are developed in Table (\ref{table_bess}). This finally leads to:
\begin{eqnarray}
<s_k^x(\valpha_p)\phi_j> &=& s_{n,m}\frac{0.023}{\pi 2^{\frac{5}{3}}\int_0^{\infty}C^2_n(h) \mathrm{d}{h}}\left(\frac{\lambda}{R_s}\right)\left(\frac{D}{r_0}\right)^{\frac{5}{3}}  \int_0^L \mathrm{d}{h}~C^2_n(h) \nonumber \\
&& \times \int_0^{\infty} \mathrm{d}\kappa~J_1\left(2\pi\zeta(h)\frac{R_s}{R}\kappa\right)J_2(2\pi\kappa)\kappa^{-\frac{11}{3}}~   \\
&& \hskip5pt \times [\beta^x_{|m|-1,k}(\theta_k^{p}(h))J_{|m|-1}\left(2\pi\rho^{p}_k(h)\kappa\right) - \beta^x_{|m|+1,k}(\theta_k^{p}(h))J_{|m|+1}(\left(2\pi\rho^{p}_k(h)\kappa\right)] \nonumber
\end{eqnarray}
Similarly, we obtain a generic expression in the $y$ direction:
\begin{eqnarray}
<s_k^y(\valpha_p)\phi_j> &=& i\frac{0.023}{2^{\frac{5}{3}}\int_0^{\infty}C^2_n(h) \mathrm{d}{h}}\left(\frac{\lambda{R}}{A_s}\right)\left(\frac{D}{r_0}\right)^{\frac{5}{3}}  \\
&&  \times  \int \mathrm{d}^2\kap  \int_0^L \mathrm{d}{h}~\widehat{\Pi_s^k}^{\ast}(\zeta(h)\kap)Q_j(\kap)  \kappa_y|\kap|^{-\frac{11}{3}} [\zeta(h)] C^2_n(h) \exp^{2i\pi\frac{h\valpha_{p}}{R}.\kap}  \nonumber
\end{eqnarray}
that assuming circular subapertures changes to:
\begin{eqnarray}
<s_k^y(\valpha_p)\phi_j> &=& s_{n,m}\frac{0.023}{\pi 2^{\frac{5}{3}}\int_0^{\infty}C^2_n(h) \mathrm{d}{h}}\left(\frac{\lambda}{R_s}\right)\left(\frac{D}{r_0}\right)^{\frac{5}{3}}  \int_0^L \mathrm{d}{h}~C^2_n(h) \nonumber \\
&& \times \int_0^{\infty} \mathrm{d}\kappa~J_1\left(2\pi\zeta(h)\frac{R_s}{R}\kappa\right)J_2(2\pi\kappa)\kappa^{-\frac{11}{3}}~   \\
&& \hskip5pt \times [\beta^y_{|m|-1,k}(\theta_k^{p}(h))J_{|m|-1}\left(2\pi\rho^{p}_k(h)\kappa\right) + \beta^y_{|m|+1,k}(\theta_k^{p}(h))J_{|m|+1}(\left(2\pi\rho^{p}_k(h)\kappa\right)] \nonumber
\end{eqnarray}
\subsection*{C.2. Computation of $<\phi^{lgs}_{1,2}(\valpha_p)\phi_{j}>$}
From Eq. (\ref{eq_ai}) we have:
\begin{eqnarray}
<\phi^{lgs}_{1}\phi_j>&=&\iint \mathrm{d}^2\ro_1\mathrm{d}^2\ro_2~\pi_p(\ro_1)Z_1(\ro_1)\pi_p(\ro_2)Z_j(\ro_2) <\phi^{lgs}(R\ro_1, \valpha_p)\phi(R\ro_2)>  \\
&=& \iint \mathrm{d}^2\ro_1\mathrm{d}^2\ro_2 \pi_p(\ro_1)Z_1(\ro_1)\pi_p(\ro_2)Z_j(\ro_2)\int^{L}_0 \mathrm{d}{h}~B_{\Delta{n}}^{h}(R[\zeta(h)\ro_1-\ro_2]+h\valpha_{p})\nonumber
\end{eqnarray}
which in the Fourier plane becomes:
\begin{eqnarray}
<\phi^{lgs}_{1}\phi_j>&=& \frac{1}{R^2}\int \mathrm{d}^2\kap \int_0^L \mathrm{d}{h}~ Q_1^{\ast}(\zeta(h)\kap)Q_j(\kap) W^{h}_{\Delta{n}}\left(\frac{\kap}{R}\right) \exp^{2i\pi\frac{h\valpha_{p}}{R}.\kap} \\
 &=& \frac{0.023}{2^{\frac{5}{3}}\int_0^{\infty}C^2_n(h) \mathrm{d}{h}}\left(\frac{D}{r_0}\right)^{\frac{5}{3}} \nonumber \\
&&\times \int \mathrm{d}^2\kap \int_0^L \mathrm{d}{h}~Q_1^{\ast}(\zeta(h)\kap)Q_j(\kap) |\kap|^{-\frac{11}{3}} C^2_n(h) \exp^{2i\pi\frac{h\valpha_{p}}{R}.\kap}
\end{eqnarray}
Switching to polar coordinates with Eq. (\ref{eq_tfzern}) we get:
\begin{eqnarray}
<\phi^{lgs}_{1}\phi_j>&=&  i(-1)^{-\frac{m}{2}}s_{n,m}\frac{4 \times 0.023}{\pi^22^{\frac{5}{3}}\int_0^{\infty}C^2_n(h) \mathrm{d}{h}}\left(\frac{D}{r_0}\right)^{\frac{5}{3}}  \int_0^L \mathrm{d}{h}~[\zeta(h)]^{-1}C^2_n(h) \nonumber \\
&&\times \int_0^{\infty} \mathrm{d}\kappa~J_2\left(2\pi\zeta(h)\kappa\right)J_{n+1}(2\pi\kappa)\kappa^{-\frac{11}{3}}  \nonumber\\
&&\times \int_0^{2\pi} \mathrm{d}\gamma \cos(\gamma)\left\{\begin{array}{c}\cos(|m|\gamma)\\\sin(|m|\gamma)\\1\end{array}\right\}\exp^{2i\pi\rho^{p}(h)\kappa\cos(\gamma-\theta^{p})} 
\end{eqnarray}
where $\rho^{p}$ and $\theta^{p}(h)$ are the modulus and the argument of $\frac{h\valpha_{p}}{R}$. We use the results of Table (\ref{table_bess}) to finally derive:
\begin{eqnarray}
<\phi^{lgs}_{1}\phi_j>&=& s_{n,m}\frac{2 \times 0.023}{\pi 2^{\frac{5}{3}}\int_0^{\infty}C^2_n(h) \mathrm{d}{h}}\left(\frac{D}{r_0}\right)^{\frac{5}{3}}  \int_0^L \mathrm{d}{h}~[\zeta(h)]^{-1}C^2_n(h) \nonumber \\
&&\times \int_0^{\infty} \mathrm{d}\kappa~J_2\left(2\pi\zeta(h)\kappa\right)J_{n+1}(2\pi\kappa)\kappa^{-\frac{14}{3}}\\
&& \hskip30pt \times [\beta^x_{|m|-1,k}(\theta^{p})J_{|m|-1}\left(2\pi\rho^{p}(h)\kappa\right) - \beta^x_{|m|+1,k}(\theta^{p})J_{|m|+1}(\left(2\pi\rho^{p}(h)\kappa\right)] \nonumber
\end{eqnarray}
The moment associated to the tilt coefficient is deduced from above by straightformward analogy:
\begin{eqnarray}
<\phi^{lgs}_{2}\phi_j>&=& \frac{0.023}{2^{\frac{5}{3}}\int_0^{\infty}C^2_n(h) \mathrm{d}{h}}\left(\frac{D}{r_0}\right)^{\frac{5}{3}} \nonumber \\
&&\times \int \mathrm{d}^2\kap \int_0^L \mathrm{d}{h}~Q_2^{\ast}(\zeta(h)\kap)Q_j(\kap) |\kap|^{-\frac{11}{3}} C^2_n(h) \exp^{2i\pi\frac{h\valpha_{p}}{R}.\kap}\\
&=& s_{n,m}\frac{2 \times 0.023}{\pi 2^{\frac{5}{3}}\int_0^{\infty}C^2_n(h) \mathrm{d}{h}}\left(\frac{D}{r_0}\right)^{\frac{5}{3}}  \int_0^L \mathrm{d}{h}~[\zeta(h)]^{-1}C^2_n(h) \nonumber \\
&&\times \int_0^{\infty} \mathrm{d}\kappa~J_2\left(2\pi\zeta(h)\kappa\right)J_{n+1}(2\pi\kappa)\kappa^{-\frac{14}{3}}\\
&& \hskip30pt \times [\beta^y_{|m|-1,k}(\theta^{p})J_{|m|-1}\left(2\pi\rho^{p}(h)\kappa\right) + \beta^x_{|m|+1,k}(\theta^{p})J_{|m|+1}(\left(2\pi\rho^{p}(h)\kappa\right)] \nonumber
\end{eqnarray}

% \bibliographystyle{osajnl}
 %\bibliography{mybib}

\begin{thebibliography}{10}
\newcommand{\enquote}[1]{``#1''}

\bibitem{foy_1}
R.~{Foy} and A.~{Labeyrie}, \enquote{{Feasibility of adaptive telescope with
  laser probe},} \aap \textbf{152}, L29--L31 (1985).

\bibitem{fugate_1}
R.~Q. {Fugate}, L.~M. {Wopat}, D.~L. {Fried}, G.~A. {Ameer}, S.~L. {Browne},
  P.~H. {Roberts}, G.~A. {Tyler}, B.~R. {Boeke}, and R.~E. {Ruane},
  \enquote{{Measurement of atmospheric wavefront distortion using scattered
  light from a laser guide-star},} \nat \textbf{353}, 144--146 (1991).

\bibitem{fried_3}
D.~L. {Fried} and J.~F. {Belsher}, \enquote{{Analysis of fundamental limits to
  artificial-guide-star adaptive-optics-system performance for astronomical
  imaging.}} Journal of the Optical Society of America A \textbf{11}, 277--287
  (1994).

\bibitem{foy_2}
M.~{Tallon} and R.~{Foy}, \enquote{{Adaptive telescope with laser probe -
  Isoplanatism and cone effect},} \aap \textbf{235}, 549--557 (1990).

\bibitem{hubin_1}
N.~{Hubin}, R.~{Arsenault}, R.~{Conzelmann}, B.~{Delabre}, M.~{Le Louarn},
  S.~{Stroebele}, and R.~{Stuik}, \enquote{{Ground Layer Adaptive Optics},}
  Comptes Rendus Physique \textbf{6}, 1099--1109 (2005).

\bibitem{fugate_2}
R.~Q. {Fugate}, B.~L. {Ellerbroek}, C.~H. {Higgins}, M.~P. {Jelonek}, W.~J.
  {Lange}, A.~C. {Slavin}, W.~J. {Wild}, D.~M. {Winker}, J.~M. {Wynia}, J.~M.
  {Spinhirne}, B.~R. {Boeke}, R.~E. {Ruane}, J.~F. {Moroney}, M.~D. {Oliker},
  D.~W. {Swindle}, and R.~A. {Cleis}, \enquote{{Two generations of
  laser-guide-star adaptive-optics experiments at the Starfire Optical Range.}}
  Journal of the Optical Society of America A \textbf{11}, 310--324 (1994).

\bibitem{lelouarn_1}
M.~{Le Louarn}, N.~{Hubin}, M.~{Sarazin}, and A.~{Tokovinin}, \enquote{{New
  challenges for adaptive optics: extremely large telescopes},} \mnras
  \textbf{317}, 535--544 (2000).

\bibitem{bonaccini_1}
D.~{Bonaccini Calia}, Y.~{Feng}, W.~{Hackenberg}, R.~{Holzl{\"o}hner},
  L.~{Taylor}, and S.~{Lewis}, \enquote{{Laser Development for Sodium Laser
  Guide Stars at ESO},} The Messenger \textbf{139}, 12--19 (2010).

\bibitem{costille_1}
A.~{Costille}, C.~{Petit}, J.-M. {Conan}, C.~{Kulcs{\'a}r}, H.-F. {Raynaud},
  and T.~{Fusco}, \enquote{{Wide field adaptive optics laboratory demonstration
  with closed-loop tomographic control},} Journal of the Optical Society of
  America A \textbf{27}, 469 (2010).

\bibitem{strobele_1}
S.~{Str{\"o}bele}, P.~{La Penna}, R.~{Arsenault}, R.~D. {Conzelmann},
  B.~{Delabre}, M.~{Duchateau}, R.~{Dorn}, E.~{Fedrigo}, N.~{Hubin},
  J.~{Quentin}, P.~{Jolley}, M.~{Kiekebusch}, J.~P. {Kirchbauer}, B.~{Klein},
  J.~{Kolb}, H.~{Kuntschner}, M.~{Le Louarn}, J.~L. {Lizon}, P.-Y. {Madec},
  L.~{Pettazzi}, C.~{Soenke}, S.~{Tordo}, J.~{Vernet}, and R.~{Muradore},
  \enquote{{GALACSI system design and analysis},} in \enquote{Society of
  Photo-Optical Instrumentation Engineers (SPIE) Conference Series,} , vol.
  8447 of \emph{Society of Photo-Optical Instrumentation Engineers (SPIE)
  Conference Series} (2012), vol. 8447 of \emph{Society of Photo-Optical
  Instrumentation Engineers (SPIE) Conference Series}.

\bibitem{conanr_1}
R.~{Conan}, F.~{Bennet}, A.~H. {Bouchez}, M.~A. {van Dam}, B.~{Espeland},
  W.~{Gardhouse}, C.~{d'Orgeville}, S.~{Parcell}, P.~{Piatrou}, I.~{Price},
  F.~{Rigaut}, G.~{Trancho}, and K.~{Uhlendorf}, \enquote{{The Giant Magellan
  Telescope laser tomography adaptive optics system},} in \enquote{Society of
  Photo-Optical Instrumentation Engineers (SPIE) Conference Series,} , vol.
  8447 of \emph{Society of Photo-Optical Instrumentation Engineers (SPIE)
  Conference Series} (2012), vol. 8447 of \emph{Society of Photo-Optical
  Instrumentation Engineers (SPIE) Conference Series}.

\bibitem{fusco_4}
T.~{Fusco}, S.~{Meimon}, Y.~{Clenet}, M.~{Cohen}, H.~{Schnetler},
  J.~{Paufique}, V.~{Michau}, J.-P. {Amans}, D.~{Gratadour}, C.~{Petit},
  C.~{Robert}, P.~{Jagourel}, E.~{Gendron}, G.~{Rousset}, J.-M. {Conan}, and
  N.~{Hubin}, \enquote{{ATLAS: the E-ELT laser tomographic adaptive optics
  system},} in \enquote{Society of Photo-Optical Instrumentation Engineers
  (SPIE) Conference Series,} , vol. 7736 of \emph{Society of Photo-Optical
  Instrumentation Engineers (SPIE) Conference Series} (2010), vol. 7736 of
  \emph{Society of Photo-Optical Instrumentation Engineers (SPIE) Conference
  Series}.

\bibitem{hu_1}
P.~H. {Hu}, J.~{Stone}, and T.~{Stanley}, \enquote{{Application of Zernike
  polynomials to atmospheric propagation problems.}} Journal of the Optical
  Society of America A \textbf{6}, 1595--1608 (1989).

\bibitem{sasiela_2}
R.~J. {Sasiela}, \emph{{Electromagnetic wave propagation in turbulence.
  Evaluation and application of Mellin transforms}} (1994).

\bibitem{sasiela_1}
R.~J. {Sasiela} and J.~D. {Shelton}, \enquote{{Mellin transform methods applied
  to integral evaluation: Taylor series and asymptotic approximations},}
  Journal of Mathematical Physics \textbf{34}, 2572--2617 (1993).

\bibitem{abramowitz_1}
M.~{Abramowitz} and I.~A. {Stegun}, \emph{{Handbook of mathematical functions :
  with formulas, graphs, and mathematical tables}} (1970).

\bibitem{gradshteyn_1}
I.~S. {Gradshteyn}, I.~M. {Ryzhik}, A.~{Jeffrey}, and D.~{Zwillinger},
  \emph{{Table of Integrals, Series, and Products}} (Seventh Edition by
  I.~S.~Gradshteyn, I.~M.~Ryzhik, Alan Jeffrey, and Daniel Zwillinger.~Elsevier
  Academic Press, 2007.~ISBN 012-373637-4, 2007).

\bibitem{tyler_1}
G.~A. {Tyler}, \enquote{{Analysis of propagation through turbulence -
  Evaluation of an integral involving the product of three Bessel functions},}
  Journal of the Optical Society of America A \textbf{7}, 1218--1223 (1990).

\bibitem{ragazzoni_3}
R.~{Ragazzoni}, S.~{Esposito}, and E.~{Marchetti}, \enquote{{Auxiliary
  telescopes for the absolute tip-tilt determination of a laser guide star},}
  \mnras \textbf{276}, L76--L78 (1995).

\bibitem{ragazzoni_2}
R.~{Ragazzoni}, \enquote{{Absolute tip-tilt determination with laser beacons.}}
  \aap \textbf{305}, L13 (1996).

\bibitem{schock_1}
M.~{Sch{\"o}ck}, R.~{Foy}, M.~{Tallon}, L.~{Noethe}, and J.-P. {Pique},
  \enquote{{Performance analysis of polychromatic laser guide stars used for
  wavefront tilt sensing},} \mnras \textbf{337}, 910--920 (2002).

\bibitem{viard_1}
E.~{Viard}, M.~{Le Louarn}, and N.~{Hubin}, \enquote{{Adaptive optics with four
  laser guide stars: correction of the cone effect in large telescopes},} \ao
  \textbf{41}, 11--20 (2002).

\bibitem{sandler_1}
D.~{Sandler}, \emph{{Laser beacon adaptive optics systems}} (1999), p. 331.

\bibitem{yura_1}
H.~T. {Yura} and M.~T. {Tavis}, \enquote{{Centroid anisoplanatism},} Journal of
  the Optical Society of America A \textbf{2}, 765--773 (1985).

\bibitem{noll_1}
R.~J. {Noll}, \enquote{{Zernike polynomials and atmospheric turbulence},}
  Journal of the Optical Society of America (1917-1983) \textbf{66}, 207--211
  (1976).

\bibitem{fried_1}
D.~L. {Fried}, \enquote{{Statistics of a Geometric Representation of Wavefront
  Distortion},} Journal of the Optical Society of America (1917-1983)
  \textbf{55}, 1427 (1965).

\bibitem{roddier_1}
F.~{Roddier}, \enquote{{The effects of atmospheric turbulence in optical
  astronomy},} Progress in optics.~Volume 19.~Amsterdam, North-Holland
  Publishing Co., 1981, p.~281-376. \textbf{19}, 281--376 (1981).

\bibitem{kolmogorov_1}
A.~{Kolmogorov}, \enquote{{The Local Structure of Turbulence in Incompressible
  Viscous Fluid for Very Large Reynolds' Numbers},} Akademiia Nauk SSSR Doklady
  \textbf{30}, 301--305 (1941).

\bibitem{tatarski_1}
V.~I. {Tatarskii}, \emph{{Wave Propagation in Turbulent Medium}} (McGraw-Hill,
  1961).

\bibitem{bufton_1}
J.~L. {Bufton}, \enquote{{Comparison of vertical profile turbulence structure
  with stellar observations.}} \ao \textbf{12}, 1785--1793 (1973).

\bibitem{greenwood_1}
D.~P. {Greenwood}, \enquote{{Bandwidth specification for adaptive optics
  systems},} Journal of the Optical Society of America (1917-1983) \textbf{67},
  390--393 (1977).

\bibitem{rousset_1}
G.~{Rousset}, \emph{{Wave-front sensors}} ({Roddier}, F., 1999), p.~91.

\bibitem{wallner_1}
E.~P. {Wallner}, \enquote{{Optimal wave-front correction using slope
  measurements},} Journal of the Optical Society of America (1917-1983)
  \textbf{73}, 1771 (1983).

\bibitem{tokovinin_1}
A.~{Tokovinin}, M.~{Le Louarn}, E.~{Viard}, N.~{Hubin}, and R.~{Conan},
  \enquote{{Optimized modal tomography in adaptive optics},} \aap \textbf{378},
  710--721 (2001).

\bibitem{bechet_1}
C.~{B{\'e}chet}, M.~{Le Louarn}, R.~{Clare}, M.~{Tallon}, I.~{Tallon-Bosc}, and
  {\'E}.~{Thi{\'e}baut}, \enquote{{Closed-loop ground layer adaptive optics
  simulations with elongated spots : impact of modeling noise correlations},}
  in \enquote{Adaptative Optics for Extremely Large Telescopes,}  (2010).

\bibitem{ragazzoni_1}
R.~{Ragazzoni}, E.~{Marchetti}, and F.~{Rigaut}, \enquote{{Modal tomography for
  adaptive optics},} \aap \textbf{342}, L53--L56 (1999).

\bibitem{schwiegerling_1}
J.~{Schwiegerling}, \enquote{{Scaling Zernike expansion coefficients to
  different pupil sizes},} Journal of the Optical Society of America A
  \textbf{19}, 1937--1945 (2002).

\bibitem{campbell_1}
C.~E. {Campbell}, \enquote{{Matrix method to find a new set of Zernike
  coefficients from an original set when the aperture radius is changed},}
  Journal of the Optical Society of America A \textbf{20}, 209--217 (2003).

\bibitem{shu_1}
H.~{Shu}, L.~{Luo}, G.~{Han}, and J.-L. {Coatrieux}, \enquote{{General method
  to derive the relationship between two sets of Zernike coefficients
  corresponding to different aperture sizes},} Journal of the Optical Society
  of America A \textbf{23}, 1960--1966 (2006).

\bibitem{bara_1}
S.~{Bar{\'a}}, J.~{Arines}, J.~{Ares}, and P.~{Prado}, \enquote{{Direct
  transformation of Zernike eye aberration coefficients between scaled,
  rotated, and/or displaced pupils},} Journal of the Optical Society of America
  A \textbf{23}, 2061--2066 (2006).

\bibitem{lundstrom_1}
L.~{Lundstr{\"o}m} and P.~{Unsbo}, \enquote{{Transformation of Zernike
  coefficients: scaled, translated, and rotated wavefronts with circular and
  elliptical pupils},} Journal of the Optical Society of America A \textbf{24},
  569--577 (2007).

\bibitem{tatulli_1}
E.~{Tatulli}, \enquote{{Transformation of Zernike coefficients: a Fourier-based
  method for scaled, translated, and rotated wavefront apertures},} Journal of
  the Optical Society of America A \textbf{30}, 726--732 (2013).

\bibitem{veran_1}
J.~P. {V{\'e}ran}, Ph.D. thesis, {\'E}cole Nationale Sup{\'e}rieure des
  T{\'e}l{\'e}communications, France, (1997) (1997).

\bibitem{fusco_1}
T.~{Fusco}, \enquote{{Optique adaptative et traitement d’images pour
  l’astronomie : de nouveaux enjeux et de nouvelles solutions },} Ph.D.
  thesis, Office national d'{\'e}tudes et de recherches a{\'e}rospatiales
  (2000).

\bibitem{fusco_2}
T.~{Fusco}, J.-M. {Conan}, V.~{Michau}, L.~M. {Mugnier}, and G.~{Rousset},
  \enquote{{Optimal phase reconstruction in large field of view: application to
  multiconjugate adaptive optics systems},} in \enquote{Society of
  Photo-Optical Instrumentation Engineers (SPIE) Conference Series,} , vol.
  4125 of \emph{Society of Photo-Optical Instrumentation Engineers (SPIE)
  Conference Series}, M.~C. {Roggemann}, ed. (2000), vol. 4125 of \emph{Society
  of Photo-Optical Instrumentation Engineers (SPIE) Conference Series}, pp.
  65--76.

\bibitem{neichel_1}
B.~{Neichel}, T.~{Fusco}, and J.-M. {Conan}, \enquote{{Tomographic
  reconstruction for wide-field adaptive optics systems: Fourier domain
  analysis and fundamental limitations},} Journal of the Optical Society of
  America A \textbf{26}, 219 (2008).

\bibitem{conan_2}
J.-M. {Conan}, Ph.D. thesis, Universit{\'e} Paris XI Orsay, (1994) (1994).

\bibitem{roddier_n_1}
N.~A. {Roddier}, \enquote{{Atmospheric wavefront simulation and Zernike
  polynomials},} in \enquote{Society of Photo-Optical Instrumentation Engineers
  (SPIE) Conference Series,} , vol. 1237 of \emph{Society of Photo-Optical
  Instrumentation Engineers (SPIE) Conference Series}, J.~B. {Breckinridge},
  ed. (1990), vol. 1237 of \emph{Society of Photo-Optical Instrumentation
  Engineers (SPIE) Conference Series}, pp. 668--679.

\bibitem{tyler_2}
G.~A. {Tyler}, \enquote{{Rapid evaluation of d$_{0}$: the effective diameter of
  a laser- guide-star adaptive-optics system.}} Journal of the Optical Society
  of America A \textbf{11}, 325--338 (1994).

\bibitem{min_1}
W.~{Min} and S.~{Yi}, \enquote{{Turbulence-induced Zernike aberrations of
  optical wavefronts in partial adaptive compensation},} Journal of Modern
  Optics \textbf{48}, 1559--1567 (2001).

\bibitem{parenti_1}
R.~R. {Parenti} and R.~J. {Sasiela}, \enquote{{Laser guide-star systems for
  astronomical applications.}} Journal of the Optical Society of America A
  \textbf{11}, 288--309 (1994).

\bibitem{wizinowich_2}
P.~{Wizinowich}, \enquote{{Progress in laser guide star adaptive optics and
  lessons learned},} in \enquote{Society of Photo-Optical Instrumentation
  Engineers (SPIE) Conference Series,} , vol. 8447 of \emph{Society of
  Photo-Optical Instrumentation Engineers (SPIE) Conference Series} (2012),
  vol. 8447 of \emph{Society of Photo-Optical Instrumentation Engineers (SPIE)
  Conference Series}.

\bibitem{kornilov_1}
V.~{Kornilov}, A.~{Tokovinin}, N.~{Shatsky}, O.~{Voziakova}, S.~{Potanin}, and
  B.~{Safonov}, \enquote{{Combined MASS-DIMM instruments for atmospheric
  turbulence studies},} \mnras \textbf{382}, 1268--1278 (2007).

\bibitem{hardy_1}
J.~W. {Hardy}, \emph{{Adaptive Optics for Astronomical Telescopes}} (1998).

\bibitem{bonaccini_2}
D.~{Bonaccini}, W.~K. {Hackenberg}, M.~J. {Cullum}, E.~{Brunetto}, T.~{Ott},
  M.~{Quattri}, E.~{Allaert}, M.~{Dimmler}, M.~{Tarenghi}, A.~{Van Kersteren},
  C.~{Di Chirico}, B.~{Buzzoni}, P.~{Gray}, R.~{Tamai}, and M.~{Tapia},
  \enquote{{ESO VLT laser guide star facility},} in \enquote{Society of
  Photo-Optical Instrumentation Engineers (SPIE) Conference Series,} , vol.
  4494 of \emph{Society of Photo-Optical Instrumentation Engineers (SPIE)
  Conference Series}, R.~K. {Tyson}, D.~{Bonaccini}, and M.~C. {Roggemann},
  eds. (2002), vol. 4494 of \emph{Society of Photo-Optical Instrumentation
  Engineers (SPIE) Conference Series}, pp. 276--289.

\bibitem{van_dam_1}
M.~A. {van Dam}, A.~H. {Bouchez}, D.~{Le Mignant}, E.~M. {Johansson}, P.~L.
  {Wizinowich}, R.~D. {Campbell}, J.~C.~Y. {Chin}, S.~K. {Hartman}, R.~E.
  {Lafon}, P.~J. {Stomski}, Jr., and D.~M. {Summers}, \enquote{{The W. M. Keck
  Observatory Laser Guide Star Adaptive Optics System: Performance
  Characterization},} \pasp \textbf{118}, 310--318 (2006).

\bibitem{max_1}
C.~E. {Max}, S.~S. {Olivier}, H.~W. {Friedman}, J.~{An}, K.~{Avicola}, B.~V.
  {Beeman}, H.~D. {Bissinger}, J.~M. {Brase}, G.~V. {Erbert}, D.~T. {Gavel},
  K.~{Kanz}, M.~C. {Liu}, B.~{Macintosh}, K.~P. {Neeb}, J.~{Patience}, and
  K.~E. {Waltjen}, \enquote{{Image Improvement from a Sodium-Layer Laser Guide
  Star Adaptive Optics System},} Science \textbf{277}, 1649--1652 (1997).

\bibitem{molodij_1}
G.~{Molodij}, \enquote{{Wavefront propagation in turbulence: an unified
  approach to the derivation of angular correlation functions},} Journal of the
  Optical Society of America A \textbf{28}, 1732 (2011).

\end{thebibliography}

%\begin{thebibliography}{99}
%\bibitem[1] . . .
%\end{thebibliography}
%\begin{thebibliography}{10}

%\end{thebibliography}

\end{document}